\documentclass[twocolumn,english,prl,longbibliography]{revtex4-1}
\usepackage{color}
\usepackage{graphicx}
\usepackage[T1]{fontenc}
\usepackage{float}
\usepackage{textcomp}
\usepackage{amsmath}
\usepackage{amssymb}
\usepackage{graphicx}
\usepackage{esint}
\usepackage{natbib}
\usepackage{color}

\begin{document}

\title{Electrical spin manipulation in double SrTiO$_3$/LaAlO$_3$ quantum dots}

\author{B. Szafran}
\email{bszafran@agh.edu.pl}
\affiliation{AGH University of Krakow, Faculty of Physics and Applied Computer Science, Al. Mickiewicza 30, 30-059 Krakow, Poland}

\author{M. Zegrodnik}
\affiliation{AGH University of Krakow, Academic Centre for Materials and Nanotechnology, Al. Mickiewicza 30, 30-059 Krakow, Poland}

\author{M. P. Nowak}
\affiliation{AGH University of Krakow, Academic Centre for Materials and Nanotechnology, Al. Mickiewicza 30, 30-059 Krakow, Poland}

\author{R. Citro}
\affiliation{Department of Physics E. R. Caianiello University of Salerno and CNR-SPIN, Via Giovanni Paolo II, 132, Fisciano (Sa), Italy}

\author{P.  W\'ojcik}
\email{pawelwojcik@agh.edu.pl}
\affiliation{AGH University of Krakow, Faculty of Physics and Applied Computer Science, Al. Mickiewicza 30, 30-059 Krakow, Poland}

\begin{abstract}
The spin dynamics in two electron double quantum dots embedded in two dimensional electron gas at the interface between SrTiO$_3$ and LaAlO$_3$ is studied by an exact numerical solution of the time-dependent Schr\"odinger equation in the context of the electric dipole spin resonance experiment. Based on the three band model of $3d$-electrons localized at Ti ions on the square lattice we analyze in details the singlet-triplet transition induced by the AC electric field, in the magnetic field range close to the avoided crossing which appears as a result of the spin-orbit coupling.
For symmetric double quantum dots the single photon spin-flip transition is prohibited due to the parity symmetry and the transition can occur only by the higher order two-photon processes. For a weakly asymmetric system, when the first order singlet-triplet transitions are released due to the parity symmetry breaking, the spin-flip transition has a character of the Rabi oscillations for a low electric field amplitude. As the amplitude is increased the frequency of the transition is blueshifted (redshifted) for the magnetic field below (above) the single-triplet avoided crossing. For a sufficiently high magnetic field and high AC field amplitude the electric field drives the system across the avoided crossing inducing the spin-flip by the Landau-Zener-Stueckelberg-Majorana effect. Finally, the optimization of the geometrical parameters of the system with respect to the time of spin-flip of its fidelity is also presented. 

\end{abstract}

\maketitle

\section{introduction}
The control the spin of electrons confined in quantum dots (QDs) has attracted a growing interest in recent years, as it lies at the heart of developments towards a scalable spin-based quantum computer~\cite{Burkard2023, Nadj-Perge2010}. The efficiency of spin manipulation is entirely contingent upon the properties of the material platform, which dictate the strength of the spin-orbit coupling (SOC)~\cite{Rashba,Rashba2003,Bychkov1984} or the hyperfine interaction with the nuclear spin bath. The latter is a source of spin decoherence~\cite{Khaetskii2002} limiting the usability of a specific material platform in accordance with one of the Loss di Vincenzo criteria~\cite{DiVincenzo1998}, whereby the quantum system must remain coherent for times much longer than the duration of elementary logic gates. For this reason, silicon-based QDs have been intensively studied over the last years due to their prolonged spin coherence time~\cite{Petta2022}.

With respect to that, another promising but still unexplored platform for QD-based spin qubits technology are transition metal oxide heterostructures, such as SrTiO$_3$/LaAlO$_3$ (LAO/STO) interfaces~\cite{Joshua2012, Maniv2015, Monteiro2019, Pavlenko2012,Biscaras2012}. The electronic properties of the two-dimensional electron gas (2DEG) formed at the LAO/STO interface is determined by the 3$d$-orbitals~\cite{Diez2015, Khalsa2013, Heeringen2013, Wojcik2021, Zegrodnik2020}, implying that direct and indirect decoherence stemming from interaction with the nuclear bath can be significantly mitigated, as it is proportional to the square of the wave function at the nuclei position~\cite{Khaetskii2002} - this value diminishes to zero for $3$d-electrons. This property, combined with the strong spin-orbit interaction~\cite{Caviglia2010,Yin2020}, high mobility~\cite{Ohtomo2004}, and susceptibility to electrostatic gates at a level comparable to semiconducting materials~\cite{Jespersen2020,Thierschmann2020,Guenevere2017}, initiates the currently ongoing research on LAO/STO-based quantum dots as potential spin-qubits with inherent scalability of 2D systems. The first experimental realization of electrostatically defined LAO/STO QDs has been already reported, yielding a Coulomb blockade diamond characteristic for a well-defined confinement~\cite{Jespersen2020,Thierschmann2020}. {\color{black} The lateral confinement within LAO/STO interface was created earlier with the litography techniques \cite{chj1} and a detection of the confined charge island, or a quantum dot in a system tailored with nanolitography has been reported \cite{chj2}. }

The demonstration of electrical spin manipulation in these systems constitutes the next milestone in spin-qubit realization based on the transition metal oxide heterostructures. Significant progress in this field has been recently achieved in electrostatically defined semiconducting QDs~\cite{Schroer2011}, where the microwave magnetic field, previously used to control spin dynamics~\cite{Koppens2006,Koppens2008}, has been replaced by the AC gate voltage~\cite{ Nadj-Perge2010, Nadj-Perge2010, Nadj-Perge2012_2, Golovach2006}.
In this technique, called electric dipole spin resonance (EDSR), the microwave potential applied to one of the gates generates a time-dependent potential which, due to the momentum-dependent spin-orbit interaction~\cite{Rashba,Rashba2003,Bychkov1984} and the resulting effective magnetic field
drives transitions between hybrydized spin-up and spin-down states, leading to Rabi oscillations~\cite{Koppens2006}. For two electrons, the SOC hybridization induces level repulsion between the spin singlet and triplet states, which is often used to determine the strength and direction of the effective SOC field~\cite{Loss2012}. Recently, EDSR technique applied to double quantum dots (DQDs) allowed for a demonstration of Pauli spin blockade~\cite{Nowack2007,Nadj-Perge2010}. In this regime, the system initially set to the spin triplet state $T_-(1,1)$, with tunneling from the left to the right QDs prohibited by selection rules, transmits to the spin singlet state $(0,2)$, which is possible due to the SOC generated by the AC field applied to one of the gates~\cite{Nowak2012}.

Although spin manipulation in LAO/STO QDs has not yet been demonstrated experimentally, theoretical studies allow us to gain a first insight into the spin dynamics in these multiorbital systems. In our recent paper~\cite{Szafran2023}, we have discussed in detail the manipulation of electron spin in a single QD in the context of EDSR experiment. We have shown that for a single electron, the spin-flip in the ground state has the character of a Rabi resonance, while for two electrons, the singlet-triplet transition is forbidden by parity symmetry. The latter one is possible via a second-order, two-photon process, which exhibits a two-state Rabi character for low AC field amplitude.

Here we extend our study to the case of two-electrons in double quantum dots. Based on the time dependent scheme we simulate the singlet-triplet transition induced by the AC electric field. We consider the magnetic field range in the nearest of the avoided crossing which appears as result of the SO coupling. The calculated transition between singlet-triplet states, its characteristics, i.e. duration time and fidelity, are discussed with respect to the system asymmetry which is needed to induce the single photon transition by the parity symmetry breaking, as well as the coupling strength between the dots. Finally, the full symmetric system is also analyzed with respect to the higher order singlet-triplet transitions. 

{\color{black} The present paper deals only with the electronic part of the system without the phonon-field.
The coupling of electrons with the charge environment via acoustic phonons, that leads to the effects of relaxation and decoherence \cite{lossreview,climente}, needs to be integrated in a future model to account for these effects. The phonon field at the LAO/STO interface is a subject of only recent studies, in particular the effects of the surface on the optical phonon branch near the energy of $\simeq 100$ meV has been determined in Ref.~\cite{phononlaosto}. }

The manuscript is organized as follows: in Sec. II we present a theoretical model of double quantum dots in 2DEG at the (001) LAO/STO interface as well as the scheme used for time-dependent two electron calculations. Sec. III contains the analysis of electronic spectrum of two-electron double quantum dots as well as results of time dependent simulations (EDSR), finally summary and conclusions are included in Sec. IV.

\section{Theoretical model}
\subsection{The single-electron problem}
We consider 2DEG in the quantum well created at the (001) LAO/STO interface \cite{Popovic2008, Pavlenko2012}.  The system is characterized by a significant atomic SO coupling and a strong Rashba interaction due to the asymmetry of the vertical electric field at the interface~\cite{Diez2015,Heeringen2013,Khalsa2013}. 
We use the real space representation \cite{Szafran2023} of the tight-binding Hamiltonian~\cite{Diez2015} spanned by $3d$ orbitals of Ti ions that are arranged in a square lattice. The Hamiltonian of the system is given by
\begin{eqnarray}
    \hat{H}&=&\sum _{\mu,\nu} \hat{C}^\dagger_{\mu, \nu} ( \hat{H}^{0}+\hat{H}_{SO}+\hat{H}_B ) \hat{C}_{\mu, \nu}+ 
    \label{eq:Hamiltonian_real} \\
    &&\sum _{\mu,\nu} \hat{C}^\dagger_{\mu+1, \nu} \hat{H}^{x} \hat{C}_{\mu, \nu}  + \sum _{\mu,\nu} \hat{C}^\dagger_{\mu, \nu+1} \hat{H}^{y} \hat{C}_{\mu, \nu} + \nonumber \\ 
    && \sum _{\mu,\nu} \hat{C}^\dagger_{\mu+1, \nu-1} \hat{H}_{mix} \hat{C}_{\mu, \nu} - \nonumber \\
    && \sum _{\mu,\nu} \hat{C}^\dagger_{\mu+1, \nu+1} \hat{H}_{mix} \hat{C}_{\mu, \nu} +h.c., \nonumber
\end{eqnarray}
where 
$\hat{C}_{\mu, \nu}=(\hat{c}_{\mu, \nu,xy}^{\uparrow}, \hat{c}_{\mu, \nu, xy}^{\downarrow}, \hat{c}_{\mu, \nu, xz}^{\uparrow}, \hat{c}_{\mu, \nu, xz}^{\downarrow}, \hat{c}_{\mu, \nu, yz}^{\uparrow}, \hat{c}_{\mu, \nu, yz}^{\downarrow})^{T}$ is the vector of electron anihilation operators corresponding to states with the spin $\sigma=\uparrow,\downarrow$ on the orbital $d_{xy}, d_{xz}, d_{yz}$ at the position $( \mu,\nu )$. In Eq. (\ref{eq:Hamiltonian_real}), $\hat{H}^{0}$ accounts for the splitting of the $3d$ orbitals degeneracy and the in-plane external potential $V(\mathbf{r})$ defining the quantum dot confinement
\begin{eqnarray}
\hat{H}^{0}&=&
\left(
\begin{array}{ccc}
 4t_l-\Delta _E & 0 & 0\\
 0 & 2t_l+2t_h & 0 \\
 0 & 0  &  2t_l+2t_h
\end{array} \right) \otimes \hat {\sigma} _0 \nonumber \\
&+&
\left(
\begin{array}{ccc}
 V_{\mu,\nu} & 0 & 0\\
 0 & V_{\mu,\nu} & 0 \\
 0 & 0  &  V_{\mu,\nu}
\end{array} \right) \otimes \hat {\sigma} _0,
\label{eq:Hamiltonian_real_H0}
\end{eqnarray}
{\color{black} where the tight-binding parameters $t_l=875\;$meV, $t_h=40\;$meV, $\Delta_E=47\;$meV, are adopted after Ref. ~\cite{Zegrodnik2020, Diez2015}, and $V_{\mu,\nu}$ is the external potential given at the ion position ($\mu,\nu$.) }

The atomic spin-orbit coupling is defined by $\hat{H}_{SO}$ and is given by
\begin{equation}
\hat{H}_{SO}= \frac{\Delta_{SO}}{3}
\left(
\begin{array}{ccc}
0 & i \sigma _x & -i \sigma _y\\
-i \sigma _x & 0 & i \sigma _z \\
i \sigma _y & -i \sigma _z & 0
\end{array} \right) \;,
\label{eq:hso}
\end{equation}
where the matrix corresponds to  ${\bf L \cdot S}$ with the orbital angular momentum represented in the $t_{2g}$-orbitals basis {\color{black} and we take the SO coupling parameter
$\Delta_{SO}=10$~meV \cite{Caviglia2010}}.

 Finally, the coupling of the external magnetic field to the spin and atomic orbital momentum of electrons is taken into account by the Hamiltonian
\begin{equation}
\hat{H}_B=\mu_B(\mathbf{L}\otimes \sigma_0+g\mathbf{1}_{3\times 3} \otimes \mathbf{S})\cdot \mathbf{B}/\hbar,
\label{eq:Hb}
\end{equation}  
{\color{black}where $g$ is the the Land\'e factor $g=3$~\cite{Ruhman2014,Jespersen2020}},
$\mu_B$ is the Bohr magneton,  $\mathbf{S}=\hbar \pmb{\sigma}/2$ with $\pmb{\sigma}=(\sigma_x,\sigma_y,\sigma_z)$ and $\mathbf{L}=(L_x,L_y,L_z)$ with
\begin{equation}
\begin{split}
 L_x&= \left ( 
 \begin{array}{ccc}
  0 & i & 0 \\
  -i & 0 & 0 \\
  0 & 0 & 0 
 \end{array}
 \right ), 
 L_y= \left ( 
 \begin{array}{ccc}
  0 & 0 & -i \\
  0 & 0 & 0 \\
  i & 0 & 0 
 \end{array}
 \right ), 
 L_z= \left ( 
 \begin{array}{ccc}
  0 & 0 & 0 \\
  0 & 0 & i \\
  0 & -i & 0 
 \end{array}
 \right ).
 \end{split}
\end{equation}
The hopping elements in Eq. (\ref{eq:Hamiltonian_real}) consist of the kinetic term (spin-conserving intersite hopping) and the Rashba SOC (mixing
between the $d_{xy}$ and the two other orbitals) and take the form
\begin{equation}
\hat{H}^{x}=
\left(
\begin{array}{ccc}
 -t_l & 0 & 0\\
 0 & -t_l & 0 \\
 0 & 0  &  -t_h
\end{array} \right) \otimes \hat {\sigma} _0
+
\frac{\Delta_{RSO}}{2}
\left(
\begin{array}{ccc}
 0 & 0 & -1 \\
 0 & 0 & 0 \\
 1 & 0  &  0
\end{array} \right) \otimes \hat {\sigma} _0\;,
\label{eq:real_space_Hx}
\end{equation}
and
\begin{equation}
\hat{H}^{y}=
\left(
\begin{array}{ccc}
 -t_l & 0 & 0\\
 0 & -t_h & 0 \\
 0 & 0  &  -t_l
\end{array} \right) \otimes \hat {\sigma} _0
+
\frac{\Delta_{RSO}}{2}
\left(
\begin{array}{ccc}
 0 & -1 & 0 \\
 1 & 0 & 0 \\
 0 & 0  &  0
\end{array} \right) \otimes \hat {\sigma} _0,
\label{eq:real_space_Hy}
\end{equation}
{\color{black}with the Rashba SO coupling parameter $\Delta_{RSO}=20$~meV \cite{Yin2020}.}
Finally, the hybridization between $d_{xz}$, $d_{yz}$ orbitals are taken into account by
\begin{equation}
\hat{H}_{mix}= \frac{t_d}{2}
\left(
\begin{array}{ccc}
 0 & 0 & 0 \\
 0 & 0 & 1 \\
 0 & 1  &  0
\end{array} \right) \otimes \hat {\sigma} _0,
\label{eq:real_space_hybrid}
\end{equation}
{\color{black} with the hybridization parameter  $t_d=40\;$meV \cite{Diez2015}.}

The external magnetic field
induces the Aharonov-Bohm phase shifts for the electrons
moving along the interface, that we account for using the Peierls approach, 
i.e. multiplying the elements of the Hamiltonian corresponding to the hopping between ions placed at points ${\bf r}_a=(x_{\mu_a},y_{\nu_a})$
and ${\bf r}_b=(x_{\mu_b},y_{\nu_b})$ by a factor $e^{{i\frac{e}{\hbar}\int_{\bf{r}_a}^{\bf{r}_b}}\vec {A}\cdot \vec {dl}}$, where ${\vec{A}}$ is the vector potential taken in the symmetric form ${\vec{A}}=(-y/2,-x/2,0)B$ for $\vec{B}=(0,0,B)$. {\color{black} Most of the results provided below are obtained
for this perpendicular magnetic field orientation. However, the results for in-plane
orientation  $\vec{B}=(B,0,0)$ are also provided. The in-plane field does not induce any Aharonov-Bohm phase shifts, produces no orbital effects and enters to the Hamiltonian only via the spin Zeeman term.}

The in-plane external potential that models the double quantum dot system is assumed as a superposition of two Gaussian functions 
\begin{eqnarray}V(x,y)&=&-V_0\exp\left(-\frac{1}{2}\left((x+\frac{s_x}{2})^2+y^2\right)/R^2\right) \nonumber \\&-&V_1\exp\left(-\frac{1}{2}\left((x-\frac{s_x}{2})^2+y^2\right)/R^2\right),
\end{eqnarray} 
where $s_x$ is the distance between the minima of the separate functions (centers of the left and right QDs), $V_0$ and $V_1$ define the depth of each potential minimum and {\color{black} $R=12$ nm} is the size parameter of the separate Gaussians.

{\color{black} In our model we assume that the potential of each of the dots separately
is rotationally symmetric. For the parameters applied below the tunnel coupling between the dots is non-negligible, so that the single-electron wave functions are extended over both the dots. Any deformation of the Gaussian potential with one of the dots will predominantly act as an effective potential shift in one of the dots. The shift would induce an asymmetry in the double dot potential that, as we show below, has a very pronounced effect on the dipole matrix elements.}

\subsection{The two-electron problem}
The two-electron problem is defined by the Hamiltonian
\begin{equation}
\hat{H}_2=\hat{H}(1)+\hat{H}(2)+\frac{e^2}{4\pi\epsilon_0\epsilon r_{12}},
\end{equation}
with  the dielectric constant $\epsilon=100$ \cite{Gariglio2015}.
{\color{black} The dielectric constant of STO can reach the values as large as 10 000
 but can be limited in an electric field. As shown in the Appendix to Ref. \cite{Szafran2023} 
 the electric field near the interface hosting 2DEG is very strong and leads to
a reduction of the $\epsilon$ value to the one we apply in the present paper.}

The problem is solved with the exact diagonalization method with the basis spanned by the atomic orbitals
\begin{eqnarray}
\Psi_q(x,y,\sigma)&=&\sum_j a_j^q d_j(x,y,\sigma)\nonumber \\ &=&\sum_{r_j,o_j,s_j} a_j^q d_{r_j,o_j}(x,y)S_{s_j}(\sigma),
\end{eqnarray}
where the summation runs over the position of ions $r_j$, orbitals $o_j$ on the ion and the $z$-component of the spin indexed by $s_j$ while
$S$ is the spin-up or spin-down eigenstate.
In the sums $j$ is equivalent to the triple of indeces $(r_j,o_j,s_j)$ and $d_{r_j,o_j}$ is one of the $3d$ orbitals localized on the ion position $r_j$.
Evaluation of the Coulomb integrals has been performed in real space using the two-center approximation \cite{Guerrero2014} and integration over the $d$ orbitals after Ref. \cite{Szafran2023}.
In the calculations, we use up to 50 lowest-energy single-electron states that produce 1225 Slater determinants as a basis for the two-electron problem.

\subsection{Time stepping}
We study the spin flips induced as singlet-triplet transitions between the eigenstates of the two-electron Hamiltonian
with an external periodic perturbation of the potential $V_{AC}(x,t)=eFx\sin(\nu t)$. 
The time evolution is determined by integration of the time-dependent Hamiltonian $\hat{H_2}(t)=\hat{H_2}+V_{AC}(x_1,t)+V_{AC}(x_2,t)$. The solution is obtained in the basis of time-independent Hamiltonian eigenstates,  $\Psi(t)=\sum_m c_m(t) \exp\left(-iE_mt/\hbar\right)| m\rangle$   with $\hat{H_2}|m\rangle=E_m|m\rangle$.  Substituting this form of the wave function to the Schr\"odinger equation and projecting the resulting equation on the eigenstates $\langle n|$ one obtains a system of equations for $c_n(t)$, 
\begin{eqnarray}i\hbar c'_n(t)=eF\sum_m&&c_m(t)\exp\left[i\left(E_n-E_m\right)t/\hbar\right] \nonumber \\ 
&&\times \sin(\nu t)\langle n|x|m\rangle,\end{eqnarray}
that we solved using the Crank-Nicolson time stepping with the two-electron ground-state as the initial condition, i.e. $c_n(t=0)=\delta_{n,1}$.
 \begin{figure}[!t]
\begin{tabular}{ll}
 \includegraphics[width=.5\columnwidth]{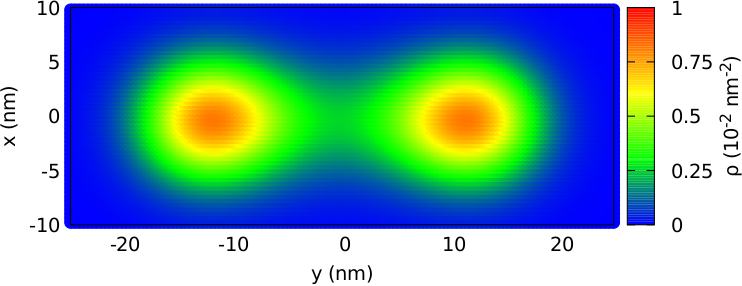} \put(-35,15){\color{white}(a)}  &
 \includegraphics[width=.5\columnwidth]{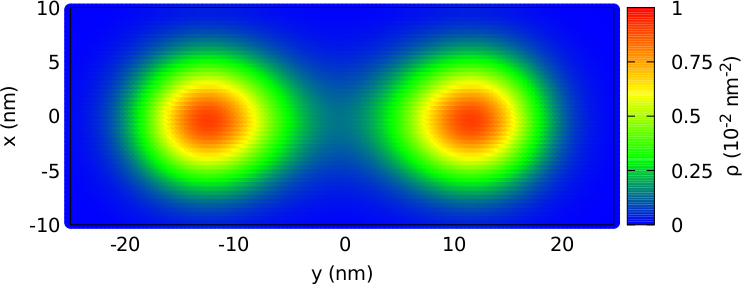} \put(-35,15){\color{white}(b)}  \\
  \includegraphics[width=.5\columnwidth]{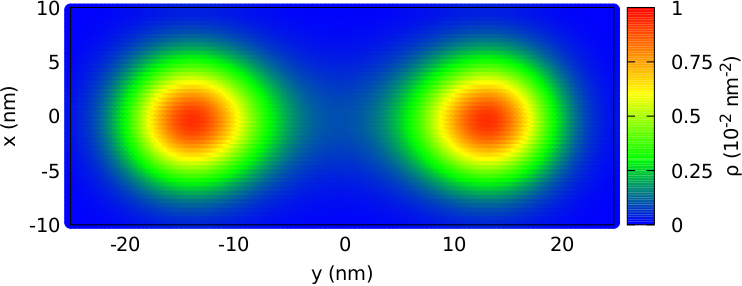} \put(-35,15){\color{white}(c)}  &
   \includegraphics[width=.5\columnwidth]{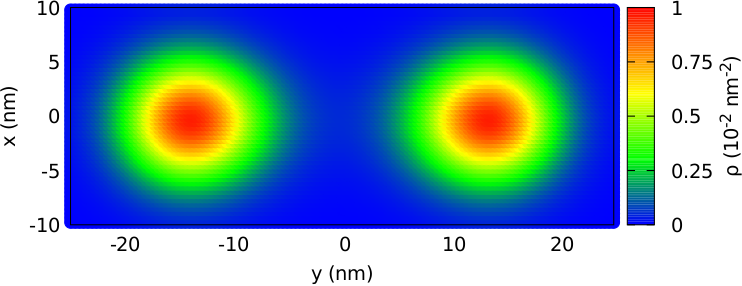} \put(-35,15){\color{white}(d)}  \\
  \end{tabular}
\caption{
The charge density for the two-electron singlet (a,c) and two-electron triplet (b,d) for $B=0$. Results for a symmetric system with $V_0=V_1=50$ meV and the spacing parameter $s_x=29.64$ nm (a,b) and $s_x=31.2$ nm (c,d).
}
 \label{ros}
\end{figure}

{\color{black}
The present model does not take into account the relaxation and dephasing
mechanisms. In a paper by two of the present authors \cite{NowakSzafran14} we
considered the effects of the relaxation via acoustic phonons for EDSR experiments 
involving two states
and demonstrated that the effect may lead to an off-resonant transition to a third
state with a lower energy than the two states coupled through a resonant AC pulse.
In Ref. \cite{Szafran2023} we argued that, when the considered transitions 
involve the two-lowest energy states in the spectrum, the relaxation mechanism is of a 
secondary importance. The dephasing in III-V and Si quantum dots is mainly due
to the hyperfine interactions with the nuclear spins which should be absent in STO
materials since the electron states are spanned by the 3d orbitals. 
The dephasing due to the interaction with the charge disorder in the surrounding 
matrix should be of a limited importance given the large value of the dielectric constant.}

\section{Results}
\subsection{The two-electron eigenstates}

In Fig. \ref{ros} we plotted the charge density of the lowest-energy spin singlet (a,c) and the lowest-energy spin triplet (b,d) for a symmetric quantum dots with 
spacing between the Gaussian minima of $s_x=29.64$~nm (a,b)  and $s_x=31.2$~nm (c,d). The results for a stronger coupling [Fig. \ref{ros}(a,b)]
and in particular the form of the densities in the barrier region 
indicate that the singlet (triplet) charge density has a bonding (anti-bonding) character. For weaker interdot coupling [Fig. \ref{ros}(c,d)] the charge densities
in both spin states are more similar and the bonding (anti-bonding) character of the states is less pronounced.
\begin{figure}
\begin{tabular}{ll}
 \includegraphics[width=.39\columnwidth]{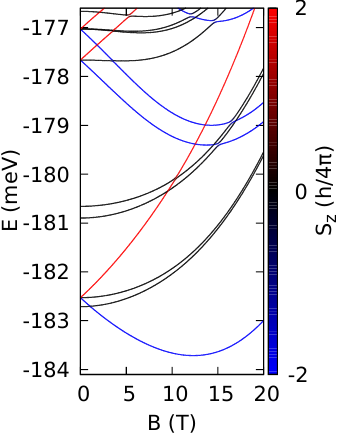} \put(-69,20){(a)}  &
 \includegraphics[width=.43\columnwidth]{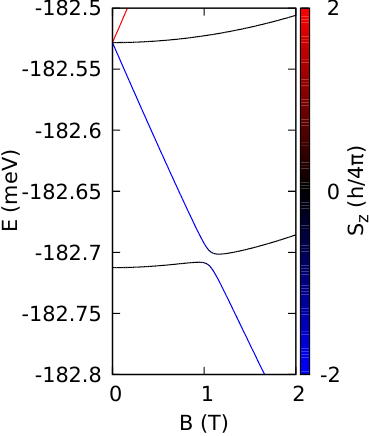} \put(-69,20){(b)}  \\
  \includegraphics[width=.39\columnwidth]{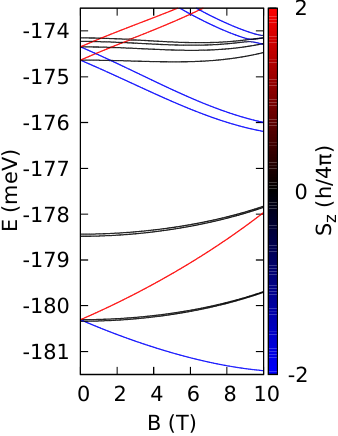} \put(-69,20){(c)}  &
   \includegraphics[width=.43\columnwidth]{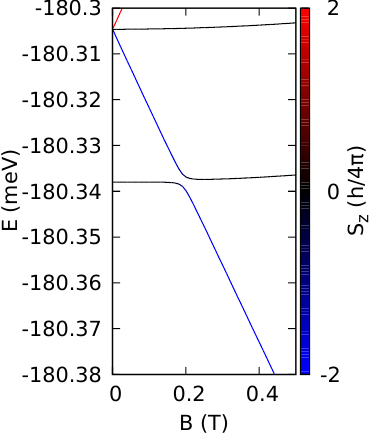} \put(-69,20){(d)}  \\
  \end{tabular}
\caption{ The low-energy spectrum for the electron pair in a symmetric
quantum dot system  $V_0=V_1=50$ meV for the spacing parameter
$s_x=29.64$~nm (a,b) and $s_x=31.2$~nm (c,d) in the perpendicular magnetic field $(0,0,B)$. The color of the line indicates
the z-component of the total spin. The singlet-triplet avoided crossing 
is enlarged in (b) and (d).
}
 \label{spesy}
\end{figure}
\begin{figure*}
    \includegraphics[width=.47\columnwidth]{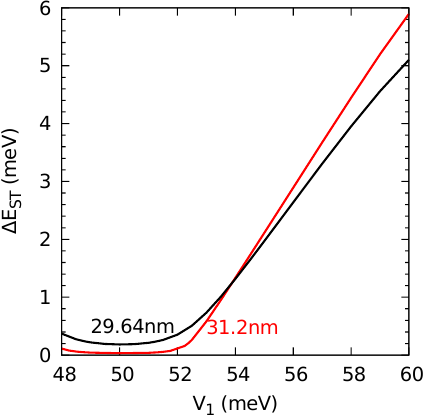} \put(-23,30){(a)}     
  \includegraphics[width=.5\columnwidth]{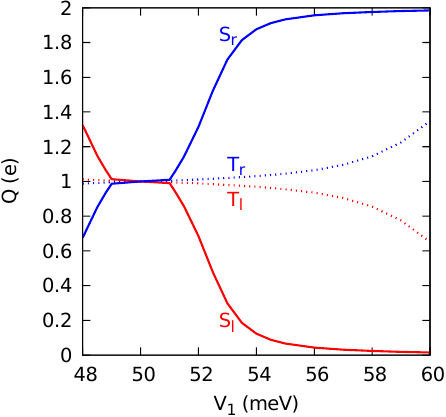} \put(-23,30){(b)} 
 \includegraphics[width=.5\columnwidth]{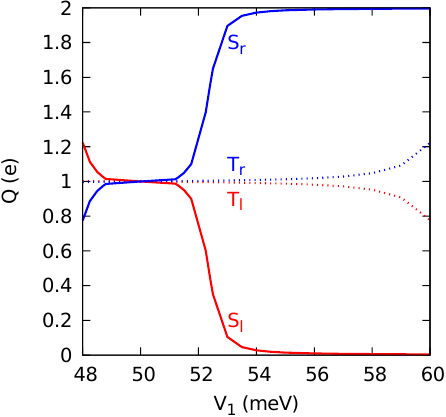} \put(-23,30){(c)}
\caption{(a) The exchange energy, i.e. the energy difference
between the spin-triplet and singlet states for $B=0$ and $s_x=29.64$ nm (black line) as well as $s_x=31.2$ nm (red line). The charge localization in the left (red line) and right dot (blue line) in the singlet (solid line) and the triplet state (dotted line)  for $s_x=29.64$nm and $s_x=31.2$ nm are displayed in (b) and (c), respectively. }
 \label{v1}
\end{figure*}

The two-electron energy spectra for the considered DQDs are displayed in Fig. \ref{spesy}(a,b) for $s_x=29.64$~nm and in Fig. \ref{spesy}(c,d) for $s_x=31.2$~nm.  At $B=0$ the ground-state is spin-singlet for any interdot distance $s_x$. The singlet-triplet energy splitting (the exchange energy) at $B=0$ varies strongly with $s_x$, e.g. the interdot tunnel coupling, and
for $s_x=29.64$~nm it is equal to 185.5~$\mu$eV, decreasing to only 33.1~$\mu$eV
for $s_x=31.2$~nm. The applied external magnetic field promotes the spin-polarized triplet for the ground-state. The singlet-triplet ground-state transformation occurs via an avoided crossing [Fig. \ref{spesy}(b,d)] that is opened
by the spin-orbit coupling. The position of the avoided crossing at the magnetic field scale is determined by the exchange energy at $B=0$. For $s_x=29.64$~nm  ($s_x=31.2$~nm) the center of the avoided crossing occurs at 1.05~T
(0.19~T) with the width (e.g. minimal
energy spacing between the anticrossing energy levels) equal to 12.3~$\mu$eV for the stronger coupling and only 2.68~$\mu$eV for the weaker coupling.

To account for an intrinsic or intentional asymmetry in the confinement potential of the double quantum dot system we introduce unequal depths of the Gaussian 
quantum wells forming the artificial molecule. In Fig. \ref{v1}(a) we plot
the exchange energy at $B=0$ as a function of the right quantum dot potential calculated for the left potential set at $V_0=50$ meV and
$s_x=29.64$nm (black line) as well as $s_x=31.2$ nm (red line).
The exchange energy is minimal for the symmetric system. For an asymmetric double dot in the spin-singlet state both the electrons tend to occupy the deeper
quantum dot [Fig. \ref{v1}(b,c)] which lowers the singlet energy with respect to the spin triplet for which the double occupancy of the lowest orbital is forbidden by the Pauli exclusion principle. In contrast, the reaction of the charge localization to the asymmetry in the triplet state is weak [Fig. \ref{v1}(b,c)].
Note, moreover, that a small asymmetry $V_1\neq V_0$ does not introduce a significant charge redistribution to the
charge in the spin-singlet state which is a result of the electron-electron interaction. Only for $V_1\leq 49$nm or $V_1\geq 51$~nm  the electrons in the singlet start to occupy the deeper dot. For the weaker interdot coupling the transition to the deeper dot is more abrupt [cf. Fig. \ref{v1}(b) and (c)].

 \begin{figure}
\begin{tabular}{l}
 \includegraphics[width=.4\columnwidth]{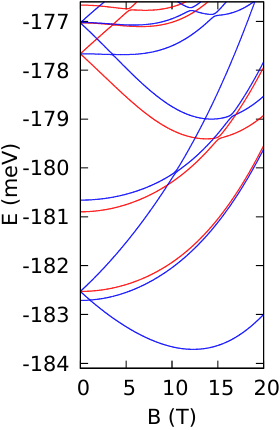} 
  \end{tabular}
\caption{Same as Fig. \ref{spesy}(a) with the color of the lines
indicates the $\Pi$ parity -- the negative one is plotted in blue,
and the positive one in red. }
 \label{par}
\end{figure}

For a symmetric potential the single-electron Hamiltonian eigenstates  are also eigenstates of the  parity operator $\Pi=\mathrm{diag}[P,-P,-P,P,-P,P]$, where $P$ is the scalar parity operator $P\psi(r)=\psi(-r)$.
The lowest-energy singlet and the lowest spin-polarized triplet that participate 
in the avoided crossing have the same -- negative -- $\Pi$ parity -- see Fig. \ref{par}. 
The same symmetry of the lowest singlet and triplet states
has a direct consequence for the spin transitions driven by AC
electric field, namely the dipole transition matrix elements between the two lowest-energy levels vanish so the first-order (single-photon) transitions is forbidden and only the second-order processes (two-photon) can occur. Similar behavior has been observed for two electrons in single quantum dot and discussed in details in our previous paper~\cite{Szafran2023}.

\subsection{EDSR in weakly asymmetric system}

Let us first study the case of weakly asymmetric system with $V_1=51$ meV.
For small asymmetry the electron charge is still more or less evenly distributed over the dots, but the asymmetry  lifts the parity selection rule that forbids the single-photon, first-order transitions.

\begin{figure}
\begin{tabular}{ll}
 \includegraphics[height=.6\columnwidth]{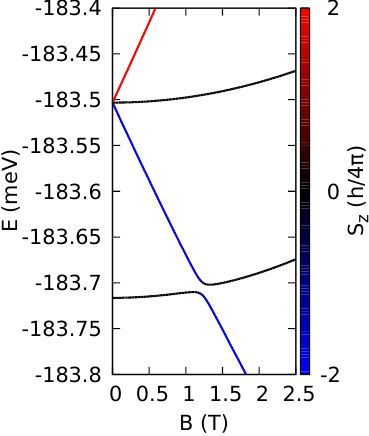}\put(-40,34){(a)}\end{tabular}
 \includegraphics[height=.5\columnwidth]{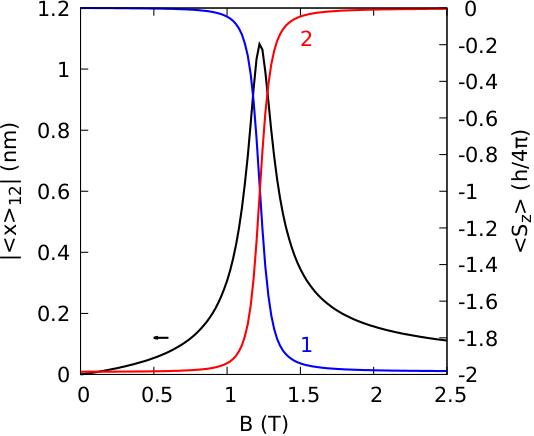}\put(-40,34){(b)} 
\caption{
(a) Same as Fig. \ref{spesy}(b) only for the right quantum dot made deeper by 1 meV,
e.g. $V_1=51$ meV. (b) The average z-component of the total spin -- with respect to the right vertical axis  -- in the ground-state (in blue) and the first excited state (in red).  The black line shows the dipole matrix element between the two lowest-energy states (referred to the left vertical axis). The numerical values of the dipole matrix elements, the energy separation from the ground-state and the value of the $z$ component of the total spin for four selected magnetic field values are given in Table I, II and III, respectively. }
 \label{asy1}
\end{figure}

\begin{table}
\begin{tabular}{ccccc}
$n$& $|x_{1n}|$(0.434T)&$|x_{1n}|$(0.94T)&$|x_{1n}|$(1.5T)&$|x_{1n}|$(2T)\\ \hline
1&2.58&2.54&0.45&0.4\\
2&0.041&0.23&0.31&0.15\\
3&$1.3\times 10^{-4}$&$7.4\times 10^{-4}$&$8.2\times 10^{-4}$&$1.1\times 10^{-3}$\\
4&$2.5\times 10^{-2}$&$4.5\times 10^{-2}$&$1.0\times 10^{-2}$&$6.1\times 10^{-3}$\\
5&$8.05$&$7.96$&$1.3$&$0.6$\\
6&$3.5$&$3.4$&$0.58$&$0.28$\\
\end{tabular}
\caption{ \color{black} The absolute value of the dipole matrix elements (in nanometers) for the six lowest energy states and the parameters
of Fig. \ref{asy1}, at four selected values of the perpendicular magnetic field $(0,0,B)$ used in Fig. \ref{asy2}.}
\end{table}

\begin{table}
\begin{tabular}{ccccc}
$n$& $\Delta E_{1n}$(0.434T)&$\Delta E_{1n}$(0.94T)&$\Delta E_{1n}$(1.5T)&$\Delta E_{1n}$(2T)\\ \hline
1&0&0&0&0\\
2&0.138&0.051&0.051&0.138\\
3&0.213&0.212&0.261&0.346\\
4&0.292&0.374&0.182&0.689\\
5&1.318&1.317&1.362&1.449\\
6&2.640&2.639&2.684&2.772\\
\end{tabular}
\caption{ \color{black}The energy spacing from the ground-state (in meV) for six lowest energy states and the parameters of Fig. \ref{asy1}.}
\end{table}

\begin{table}
\begin{tabular}{ccccc}
$n$& $S_z$(0.434T)&$S_z$(0.94T)&$S_z$(1.5T)&$S_z$(2T)\\ \hline
1&$-1.4\times 10^{-3}$ &$-1.9\times 10^{-2}$&-1.94&-1.98\\
2&-1.99&-1.96&$-4.6\times 10^{-2}$&$-6.5\times 10^{-3}$\\
3& $-5.06\times 10^{-7}$&$2.85\times 10^{-7}$&$1.14\times 10^{-6}$&$1.87\times 10^{-6}$\\
4&1.985&1.985&1.985&1.985\\
5&$1.2\times 10^{-4}$&$2.7\times 10^{-4}$&$4.5\times 10^{-4}$&$6.6\times 10^{-4}$\\
6&$-1.7\times 10^{-5}$&$-3.7\times 10^{-5}$&$-6\times 10^{-5}$&$-8.2\times 10^{-5}$\\
\end{tabular}
\caption{\color{black} The $S_z$ component of the spin in $\hbar/2$ units for the six lowest energy states and the parameters of Fig. \ref{asy1}. Results for magnetic fields $(0,0,B)$ that are used in Fig. 6.}
\end{table}

Fig. \ref{asy1}(a) shows the spectrum for $s_x=29.64$~nm near the singlet-triplet avoided crossing that is shifted to higher magnetic field by the asymmetry due to the increased exchange energy [Fig. \ref{v1}(a)] (to $B=1.22$~T from $1.05$~T in the $V_0=V_1$ case).
The width of the avoided crossing is also enlarged: from 12.3$~\mu$eV (for $V_1=50$ meV) to 13.8~$\mu$eV (for $V_1=51$~meV). The essential quantity for the first-order spin transitions is the dipole matrix element $x_{1n}=\langle \psi_1 |x| \psi_n \rangle$ between the two-lowest energy levels which is plotted by the black curve in Fig. \ref{asy1}(b) and listed in Table I for selected values of the magnetic field. In Table I the first row corresponds to the average value of $x$ for the ground-state which are about 2.5 nm due to the shift of the wave function to the deeper dot. The shift gets smaller after the avoided crossing when the spin-polarized triplet replaces the spin-singlet in the ground-state. By Pauli exclusion principle the electrons in the triplet state cannot contain an admixture of configuration with both electrons in the single-electron ground-state of the deeper dot, hence the smaller shift to the deeper quantum dot for the triplet state.
For an ideally symmetric system $V_1=V_0$ a similar singlet-triplet avoided crossing 
of energy levels is obtained with the interchange of the average values
of the spin between the two lowest-energy levels, but the transition dipole matrix element is zero. For the system with lifted parity symmetry, the matrix element reaches its maximum in the center of the avoided crossing which allows the first order transition possible.

\begin{figure}
 \begin{tabular}{llll}
\includegraphics[trim=80 0 30 0,clip,height=.5\columnwidth]{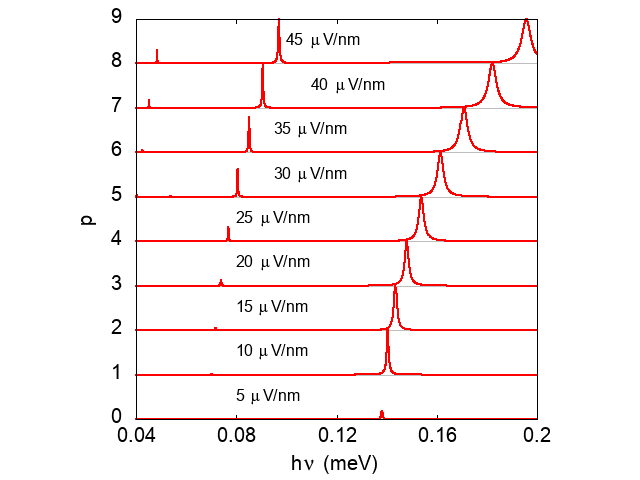}\put(-120,122){(a)} &
 \includegraphics[trim=80 0 30 0,clip,height=.5\columnwidth]{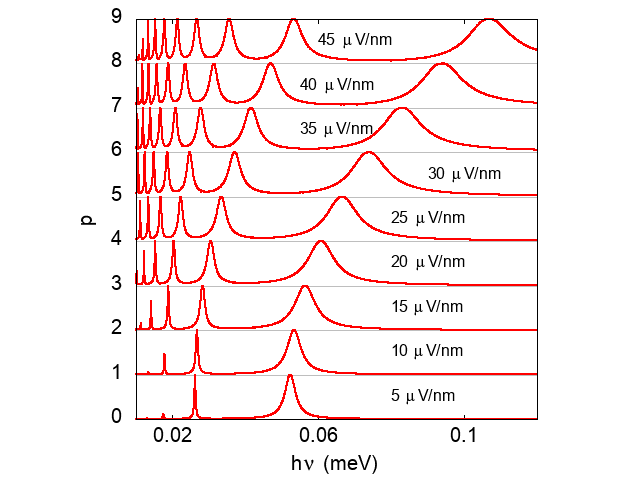}\put(-120,122){(b)}  \\
 \includegraphics[trim=80 0 30 0,clip,height=.5\columnwidth]{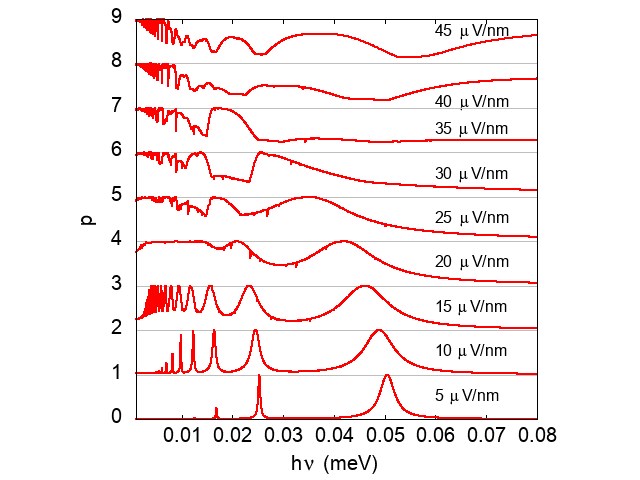}\put(-120,122){(c)} &
 \includegraphics[trim=80 0 30 0,clip,height=.5\columnwidth]{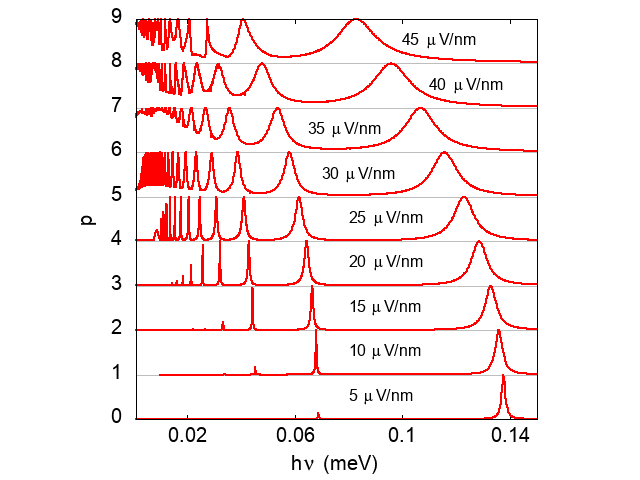}\put(-120,122){(d)}   
\end{tabular} 
\caption{
The results of simulation of the time evolution for the weakly asymmetric 
($V_1=51$~meV) double quantum dot system with spacing of $s_x=29.64$~nm
in an AC external electric field $eFx\sin(\nu t)$. 
The subsequent lines in the upper part of the plots correspond to the AC field amplitude
increased by 5~$\mu$V/nm offset by +1 each along the vertical axis.
The simulation lasts $5$~ns and starts from the ground-state. The lines show the maximal occupancy of the first excited state.
The panels correspond to $B=0.434$~T (a), $B=0.94$~T (b), $B=1.5$~T (c) and $B=2$~T (d). 
}
 \label{asy2}
\end{figure}

We studied the transitions from the ground-state to the first-excited state near the single-triplet avoided crossing induced by the external AC sine electric field, with the initial state set to the ground-state.  The maximal occupancy of the first-excited state for the simulation lasting $5$~ns is given in Fig. \ref{asy2}
for $B=0.434$~T (a), $B=0.94$~T (b), $B=1.5$~T (c), and $B=2$~T (d) with the amplitude of the AC field increasing in steps of 5~$\mu$V/nm. 

\begin{figure}
 \begin{tabular}{ll}
\includegraphics[width=.45\columnwidth]{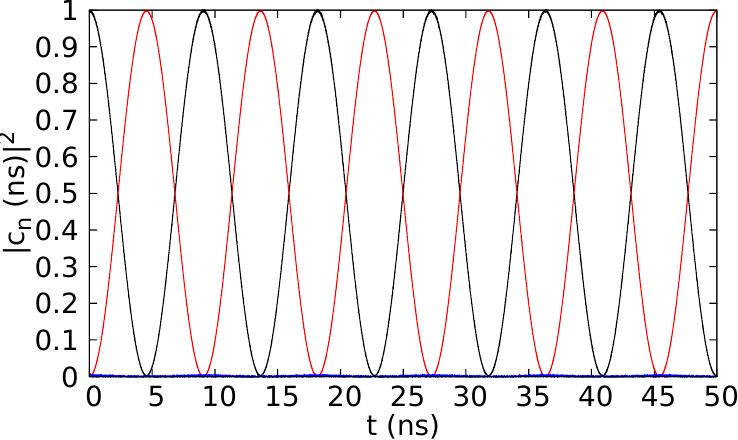}\put(-15,12){(a)} &
 \includegraphics[width=.45\columnwidth]{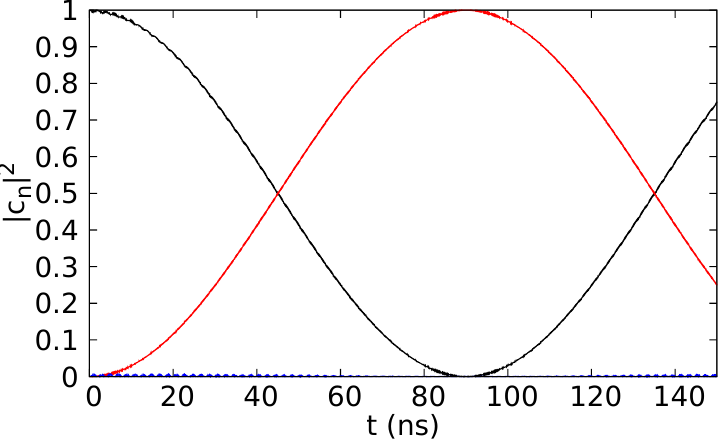}\put(-15,12){(b)}  \\
 \includegraphics[width=.45\columnwidth]{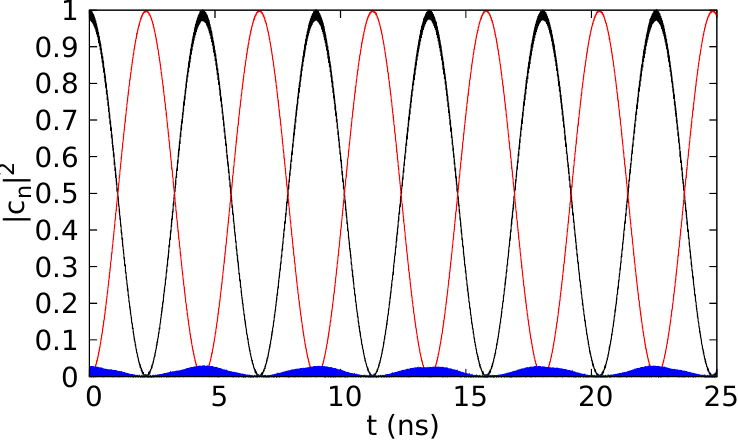}\put(-15,12){(c)} &
 \includegraphics[width=.45\columnwidth]{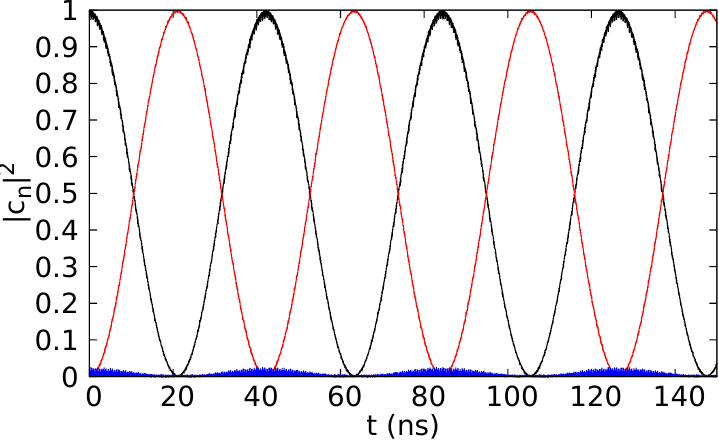}\put(-15,12){(d)}  \\
 \includegraphics[width=.45\columnwidth]{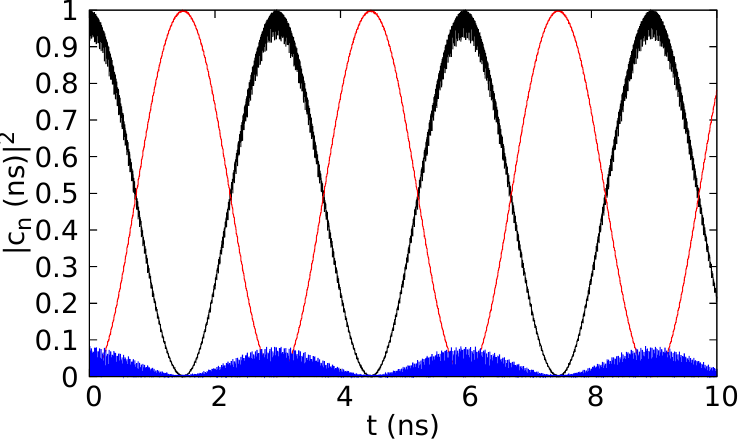}\put(-15,12){(e)} &
 \includegraphics[width=.45\columnwidth]{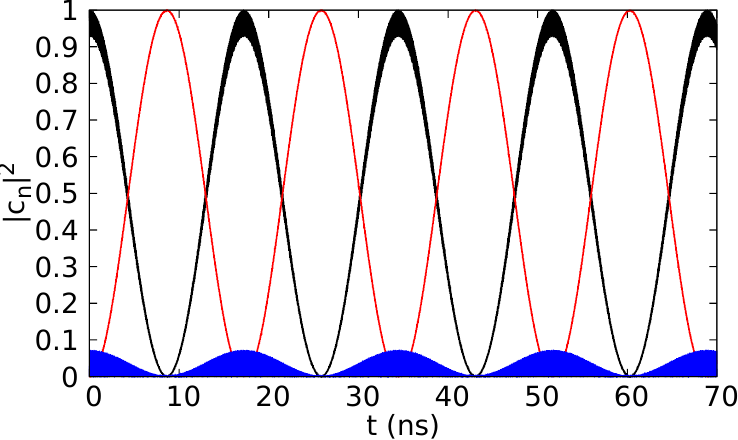}\put(-15,12){(f)}  \\
 \includegraphics[width=.45\columnwidth]{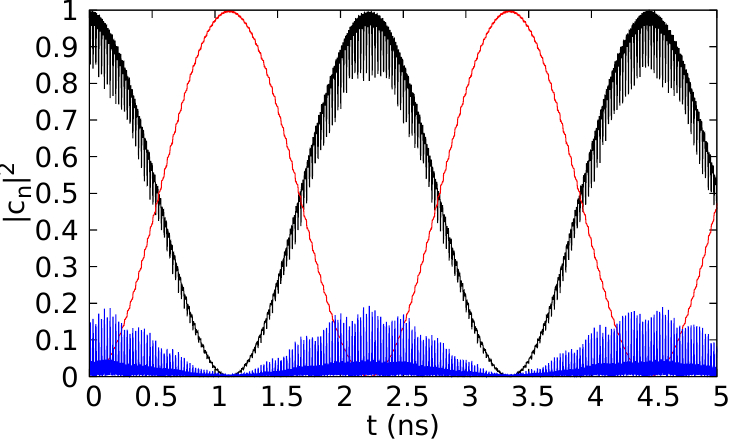}\put(-15,12){(g)} &
 \includegraphics[width=.45\columnwidth]{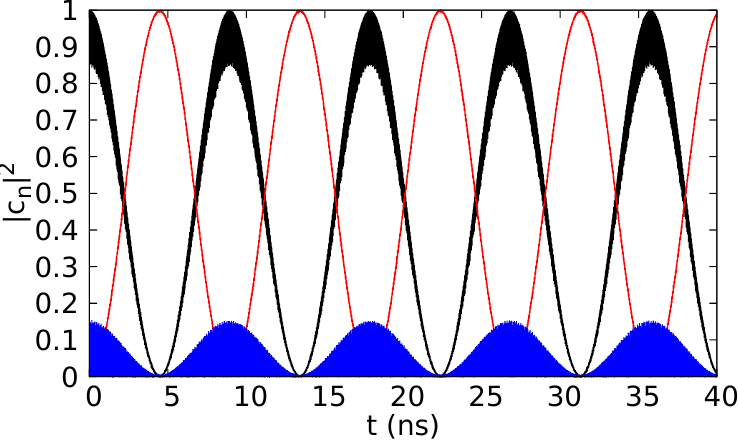}\put(-15,12){(h)}  \\

\end{tabular} 
\caption{
\color{black}
The occupation of the lowest-energy singlet (black line) and the lowest spin-polarized triplet (red line) 
as a function of time for parameters of Fig. \ref{asy2}(a), e.g. $B=0.434$ T, and the AC amplitude
of $10\mu$eV (a,b), $20\mu$eV (c,d),  $30\mu$eV (e,f),  $40\mu$eV (g,h).
The blue line shows the overall contribution from the higher excited states. 
In the left column of the plots we take the resonant frequency corresponding to the single-photon
i.e. direct or single-photon singlet-triplet transition -- the first peak on the right in Fig. \ref{asy2}(a). 
In the right column of the plots we take the frequency corresponding to the subharmonic, second-order, i.e. two-photon transition that correspond to the second peak from the right in Fig.\ref{asy2}(a). The AC frequency for the second-order transition is roughly equal to half
of the first order transition frequency.
 \label{timesy1}}
\end{figure}

\begin{figure}
\includegraphics[width=.65\columnwidth]{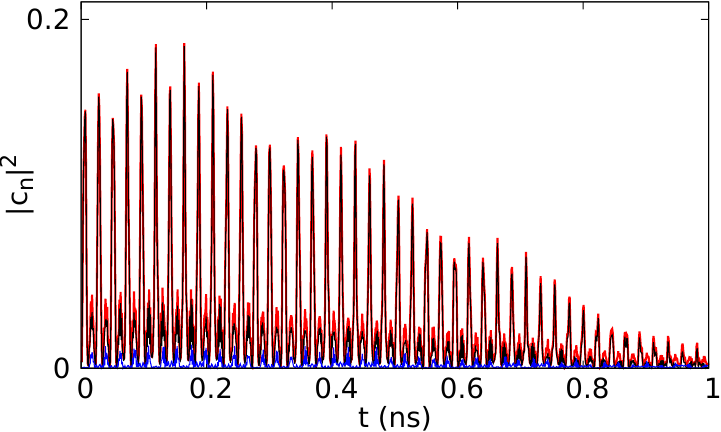}
\caption{\color{black}The overall contribution of higher energy states (red line) beyond the singlet-triplet qubit for the time evolution of Fig. \ref{timesy1}(g) and the contribution of the 5th (black line) and 6th (blue line) excited singlet states.}
 \label{timesy2}
\end{figure}

{\color{black} Fig. \ref{asy2} shows the maximal probability of the singlet-triplet transitions
that are found during 5 ns simulation for a range of AC amplitudes and four selected
perpendicular magnetic fields 0.434 T, 0.94 T, 1.5 T and 2 T. The matrix dipole elements for the singlet-triplet transitions, the energy above the ground-state  and the z-component of the total spin
for the four selected magnetic field values are given in Table I, II and III, respectively. The values
are selected symmetrically with respect to the singlet-triplet avoided crossing -- see second 
row of Table II.}
For the magnetic fields in Fig. \ref{asy2}(a,b) and Fig. \ref{asy2}(c,d) the ground-state is singlet and triplet, respectively. 
At the amplitude of 5~$\mu$eV the largest peak corresponds to the singlet-triplet transition that occurs at the driving energy near the singlet-triplet energy difference. For $B=0.94$~T and $1.5$~T, where the dipole matrix element is large, at the lowest-amplitude,
we also resolve the peaks of higher-order transitions at half and $1/3$rd of the frequency for the single-photon transition processes. 
{\color{black} In an ideal symmetric system that we considered previously \cite{Szafran2023} the
first-order singlet-triplet transitions are forbidden by the $\Pi$ parity symmetry reasons.
The transition matrix element vanish for the states that have the same parity 
as in the case of the singlet and the spin-polarized triplet in an inversion-symmetric system (cf. Fig. \ref{par}). The singlet-triplet
transitions can still be induced for a symmetric system but with a subharmonic, two-photon
transitions. The two-photon transitions at low AC frequency appear at the energy equal to roughly half the energy difference between the singlet and triplet states. The multiphoton transitions
are higher order processes that require a much longer spin-flip time than the single-photon transitions. }
As the AC field amplitude is increased the frequency of the
spin-flip transition [Fig. \ref{asy2}(a,b)] for the magnetic field below the single-triplet avoided crossing  is blueshifted at higher AC field amplitudes. On the other hand the frequency for the spin transition for the magnetic field above the single-triplet avoided crossing gets redshifted for high AC amplitudes [Fig. \ref{asy2}(c,d)]. In both cases, in an intense AC field, we find an effective upbeat of the triplet energy with respect to the singlet. 

{\color{black} Fig. \ref{timesy1} shows the time-resolved occupation of the ground-state singlet
(black line) and the excited triplet (red line) as well as the overal occupation of higher
excited states (blue) line for $B_z=0.434$ T for the amplitude of the AC field increasing from 
10$\mu$eV/nm (the highest row of plots) to 40$\mu$eV/nm (the lowest row of plots). 
The evolution shows the results corresponding to Fig. \ref{asy2}(a), only
longer evolution times than 5ns were taken when necessary. The left column of the plots in Fig. \ref{timesy1} 
shows the time evolution for the resonant single-photon frequency and the right column of the plots
to the first subharmonic (two-photon) transition. The AC frequencies corresponds to the rightmost
and second to the right peaks in Fig. \ref{asy2}(a). 
 Note the difference of the times necessary to flip the spins in the single- and two-photon-transitions (left vs right column of the plots in Fig. \ref{timesy1}). 
 The blue line in Fig. \ref{timesy1} corresponds to the overall contributions of
 the higher energy-states. This contribution is nearly missing for the lowest amplitude
 considered in Fig. \ref{timesy1}(a,b), and grows for higher AC amplitudes. 
 Note, that the triplet occupation (the red line in Fig. \ref{timesy1}) preserves its
 sine form also for higher AC amplitudes. On the other hand there is a fast oscillation
 that is superposed on the cosine form of the occupation line for the ground-state singlet.
 The ground-state singlet is strongly dipole-coupled to the 5th and 6th singlet states (see Table I)
 that are much higher in energy (Table II) with the large energy difference leading to 
 a fast rate of these off-resonant transitions between the ground-state and excited singlets. 
 The contribution of the higher excited states vanishes when the occupation of the ground-state
 singlet is zero. 
Figure \ref{timesy2} shows the overall contribution for all the excited states (black line) in the time
evolution of Fig. \ref{timesy1}(g) and the separate contributions from the 5-th and 6-th states 
(black line and blue line, respectively). The contribution of the 5-th state is larger 
consistently with the values of the dipole matrix element of Table I (see the first column
of Table I
corresponding to the perpendicular magnetic field of 0.434 T). Contribution of higher 
energy states is negligible. 
 }

\begin{figure}
 \begin{tabular}{c}
 \includegraphics[height=.45\columnwidth] 
 {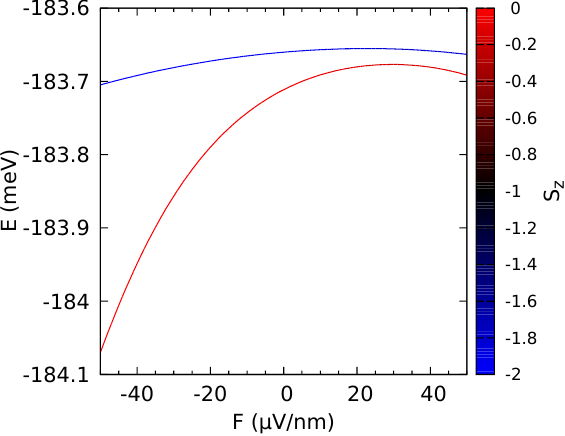} \put(-40,24){(a)} \\ \includegraphics[height=.45\columnwidth]{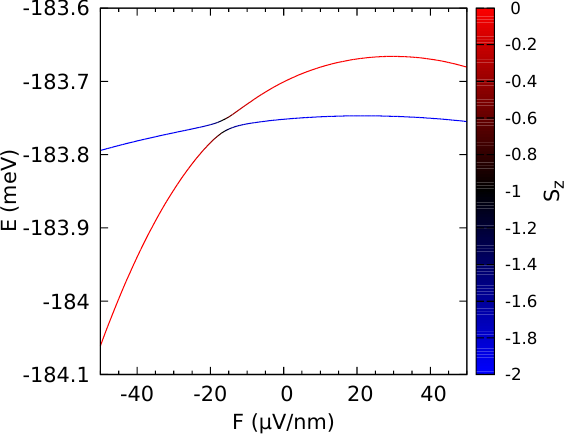}\put(-40,24){(b)}  \\ \includegraphics[height=.45\columnwidth]{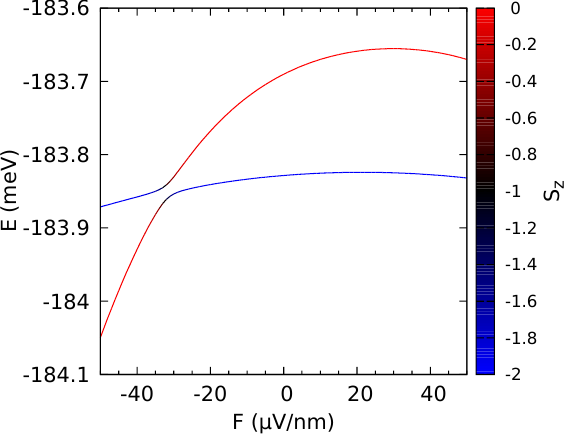}\put(-40,24){(c)} 
 \end{tabular}
\caption{
The lowest singlet and triplet energy levels for $B=0.94$~T (a), $1.5$~T (b) and $2$~T (c) for the weakly asymmetric system and a constant electric field $+eFx$. In the considered field range a singlet-triplet avoided crossing opens for $B=1.5$~T and $2$~T that changes.
In the AC simulation the avoided crossing changes the spin-flip probability
at low frequency from the multiphoton transitions to the LZSM interference -- see Fig. \ref{asy2}(c,d) and Fig. \ref{asy3} at higher amplitude.
} \label{szaxc}
 \end{figure}

{\color{black} The spectrum for the magnetic field increased from 0.434T to 0.94T [Fig. \ref{asy2}(b)]
has much better resolved subharmonic structure  due to $\simeq 5.6$ times
larger dipole matrix element (the second row of Table I and Fig. \ref{asy1}(b)).
In Fig. \ref{asy2}(c) for 1.5T we find that the regular transition spectrum gets deteriorated at higher frequency with the disappearance of the single-photon peak. A near structure of peaks
is restored at 2T -- see Fig. \ref{asy2}(d). Both Fig. \ref{asy2}(c,d) in contrast to 
Fig. \ref{asy2}(a,b) develop a large values of spin-flip probability at the lowest
range of the frequency. This is due to the electric field sweeping across
the singlet-triplet avoided crossing. In order to explain this effect in
Fig. \ref{szaxc} we show} the two lowest-energy states for the double dot system with a time-independent electric field of $eFx$ added to the potential.
The avoided crossing for $B=1.5$~T [Fig. \ref{szaxc}(b)] is reached at $|F|\simeq 20~\mu$V. Note, that in AC field with the amplitude of $F\simeq 20~\mu$V/nm the
maximal occupancy of the singlet-state is no longer zero at low $h\nu$ limit [cf. Fig. \ref{asy2}(c)]. 
At higher amplitude the low-frequency limit in Fig. \ref{asy2}(c) corresponds to an adiabatic sweep of the potential across the singlet-triplet avoided crossing with 100\% spin flip probability in a limit of an adiabatic sweep. 

\begin{figure}
\includegraphics[trim=30 0 30 0,clip,height=.65\columnwidth]{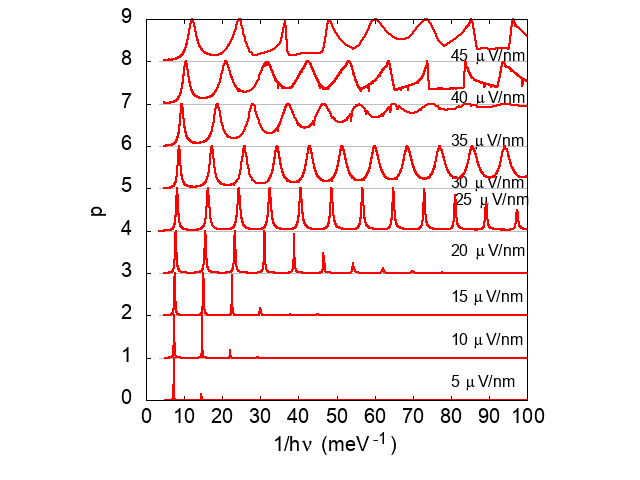} \caption
{Same as Fig. \ref{asy2}(d) only as a function of the inverse of the driving energy.}
 \label{asy3}
\end{figure}

{\color{black} For $B=1.5$~T and $B=2$ T at higher amplitude the AC field drives 
the system across the singlet-triplet avoided crossing so that the process for the spin-flip  can acquire the Landau-Zener-Stueckelberg-Majorana (LZSM) character \cite{lzs}.
The ground-state system  that is driven across the avoided crossing can either stay in the ground-state (adiabatic limit) with the transformation of the wave function from the one 
characteristic to the ground-state at the other side of the avoided crossing or preserve its wave function and find itself in the excited state. For periodic modulation the phases for 
both processes accumulate or cancel that leads to the LZSM interference pattern \cite{lzs}.}
The interference produces the comb visible in Fig. \ref{asy2}(c) at higher amplitude
and low frequency. For $B=2$~T we reach a similar point at $F\simeq 30~\mu$V/nm [cf. Fig. \ref{asy2}(d) and Fig. \ref{szaxc}(c)].
The change of the character of the triplet-singlet transition is better illustrated
by the transfer probability plotted as a inverse of the driving energy
presented in Fig. \ref{asy3} for $B=2$~T.
As the amplitude increases from 5$~\mu$V/nm to 25$~\mu$V/nm 
higher-order peaks corresponding to multiphoton transitions appear
but the corresponding peaks fall below 100\% at large $1/h\nu$.
For $30~\mu$V/nm and above, the transition probability peaks acquire the height of 100\%  that does not fall below at large $1/h\nu$, and appear periodically on the $1/h\nu$ scale, which are  characteristics of the LZSM interference \cite{lzs}.

\begin{figure}
 \begin{tabular}{ll}
\includegraphics[width=.45\columnwidth]{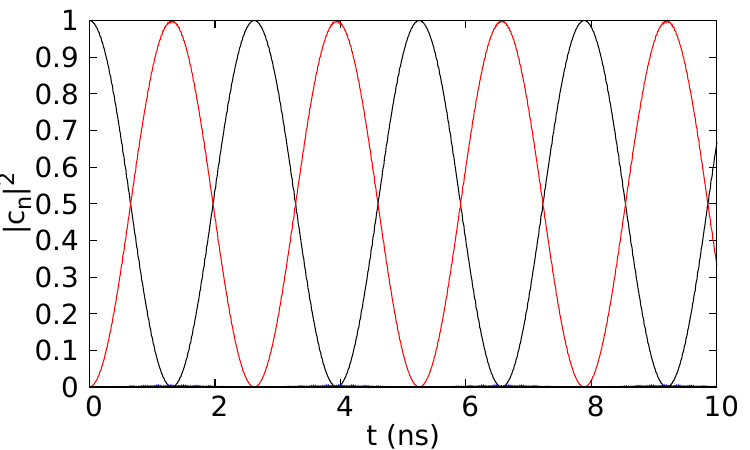}\put(-15,12){(a)} &
 \includegraphics[width=.45\columnwidth]{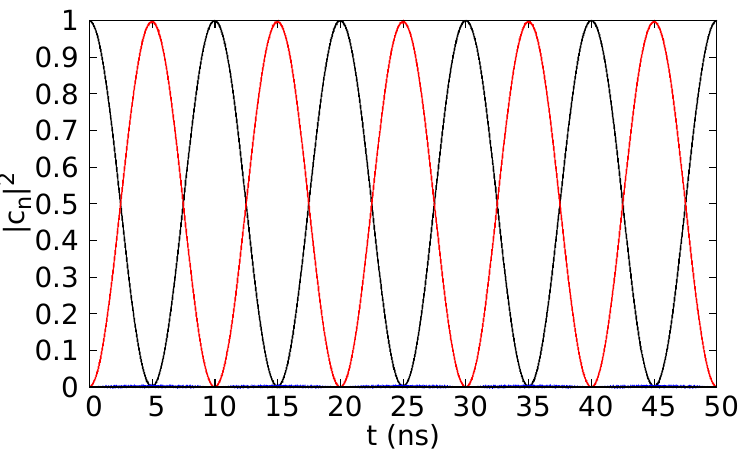}\put(-15,12){(b)}  \\
\includegraphics[width=.45\columnwidth]{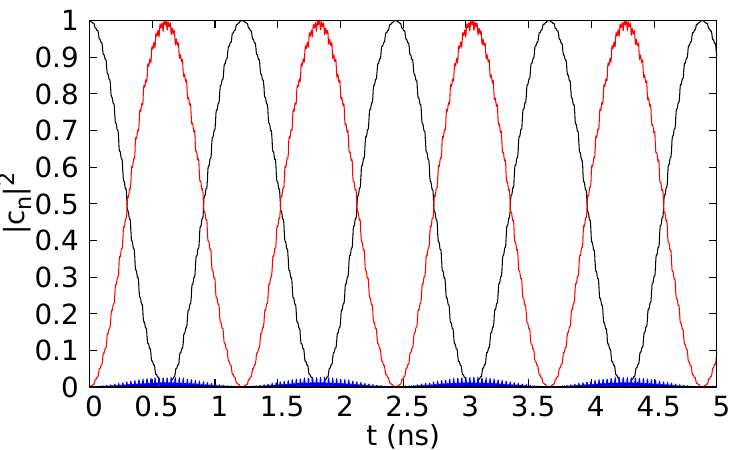}\put(-15,12){(c)} &
 \includegraphics[width=.45\columnwidth]{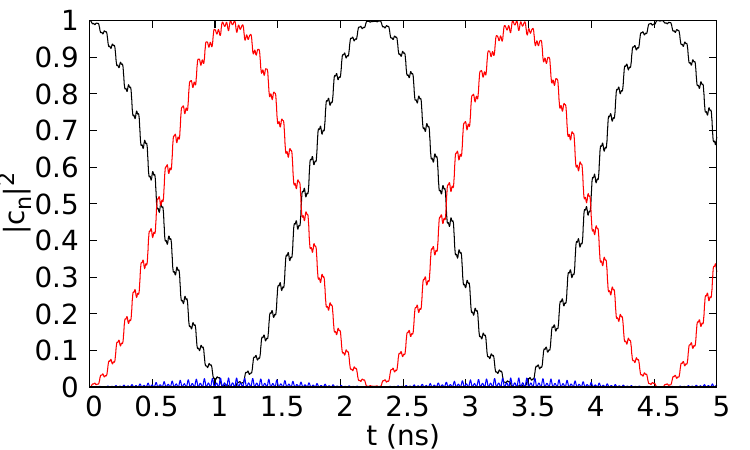}\put(-15,12){(d)}  \\
\includegraphics[width=.45\columnwidth]{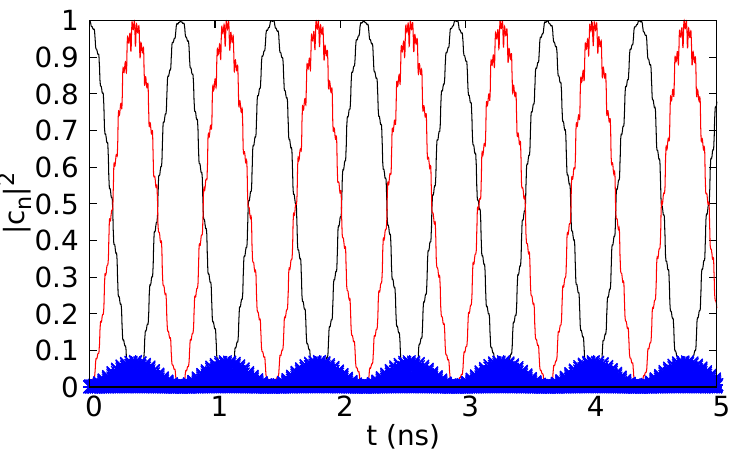}\put(-15,12){(e)} &
 \includegraphics[width=.45\columnwidth]{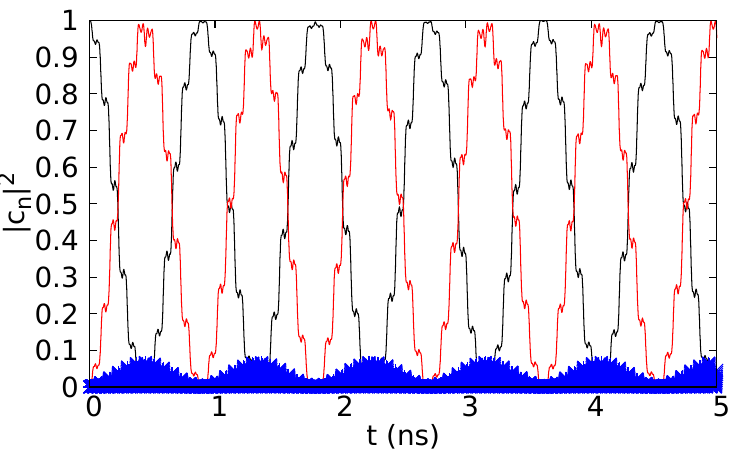}\put(-15,12){(f)}  \\
\includegraphics[width=.45\columnwidth]{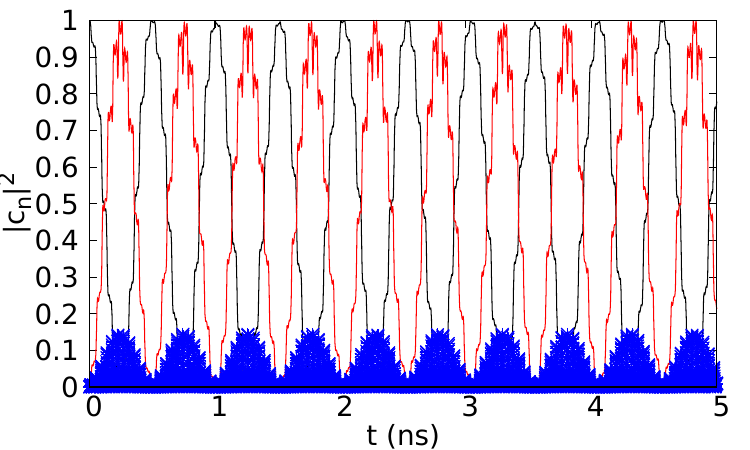}\put(-15,12){(g)} &
 \includegraphics[width=.45\columnwidth]{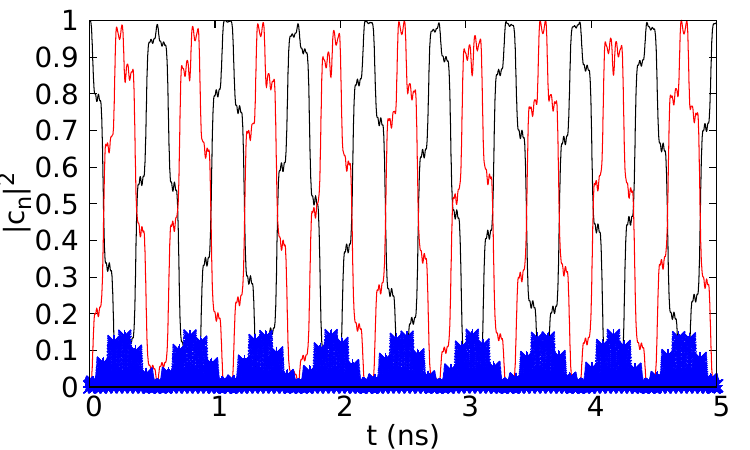}\put(-15,12){(h)}  
\end{tabular} 
\caption{\color{black}
Same as Fig. \ref{timesy2} only for $B=2T$.
}
 \label{timesy2x}
\end{figure}

\begin{figure}
\begin{tabular}{l}
\includegraphics[width=.85\columnwidth]{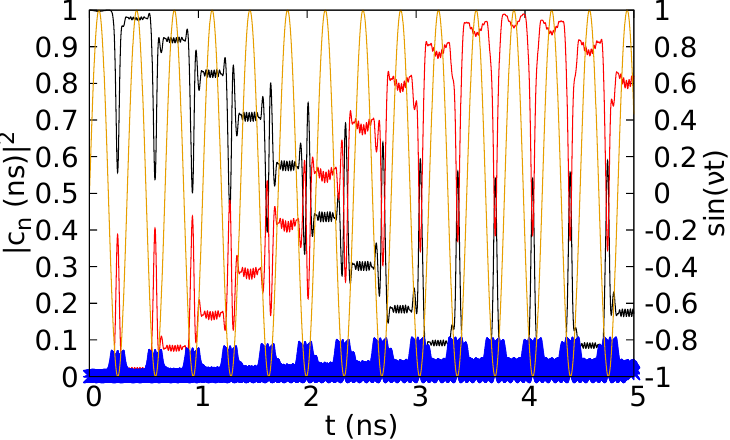}\\\includegraphics[width=.85\columnwidth]{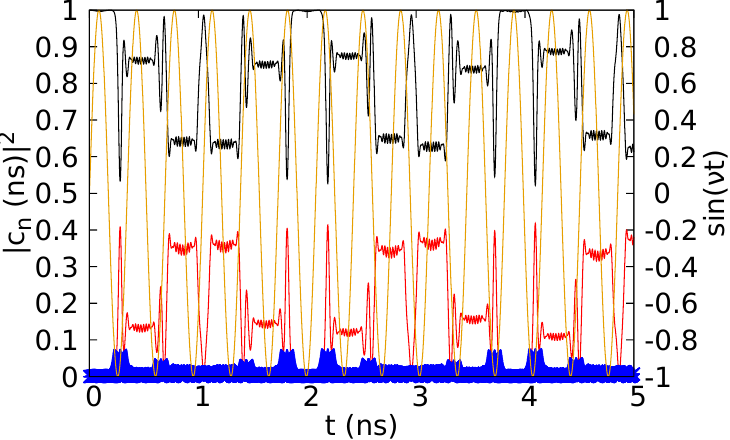}
\end{tabular}
\caption{The time-resolved occupation of the triplet ground-state (black line) and the 
lowest singlet-state (red) line for $B=2$T with parameters of Fig. 10 and Fig. 6(d)
for AC amplitude of 40$\mu$V/nm with $h\nu=0.0119295$ meV (a) and $h\nu=0.0108441$ meV (b).
Panel (a) corresponds to the last peak at Fig. \ref{asy3} and
panel (b) to a minimum between the two last peaks. In orange we plotted the $\sin\nu t$ function
that drives the system across the avoided crossing shown in Fig. 9(c). }
 \label{lzt}
\end{figure}

{\color{black} The time-resolved occupation of the states involved in the system dynamics
for the magnetic field of 2T is illustrated in Figs. \ref{timesy2x} and \ref{lzt}.
Fig. \ref{timesy2x} corresponds to the one- and two-photon transitions.
At the AC amplitude of $10\mu$V/nm [Fig. \ref{timesy2x}(a,b)] the occupation has a 
similar character to the ones found for $B=0.434$ T [cf. \ref{timesy1}(a,b)].
At higher amplitude the contribution from the higher excited singlet states (blue line in Fig. \ref{timesy2x}(e-h))
is correlated with the red line -- that now corresponds to the spin-singlet which
is the first excited state in the field of 2T. 
For the amplitude of 20$\mu$V/nm the system is still not driven across the avoided 
crossing, only approach it (see Fig. \ref{szaxc}(c)). 
When this happens (cf. Fig. \ref{timesy2x}(e,f) and (g,h) for the amplitude of the AC
field of 30 and 40$\mu$V/nm, respectively) the time evolution of the occupations
deviates visibly from the simplest periodicity -- see the shapes of the maximal occupation probability at each turn.  Note, that the time evolution in Fig. \ref{timesy1} for the magnetic field of 0.434 T, when the sweep across the avoided crossing does not occur, is periodic in the entire range of frequencies considered.  In Fig. \ref{timesy2x} for higher AC amplitudes the second order transition becomes nearly as fast as for the 
two-photon transition which never occurs for the driving off the avoided crossing.
The driving in Fig. \ref{timesy2x} corresponds to the first two peaks at right of Fig. \ref{asy2}(d), i.e. to a relatively large AC frequency. The time evolution for the lowest-frequency
i.e., rightmost peak in Fig. \ref{asy3} is given in Fig. \ref{lzt}(a). The orange line in this peak shows
the $\sin\nu t$ curve that modulates the external field. The center of the AC 
of Fig. \ref{szaxc}(c) is reached at $\sin\nu t\simeq -0.8$. Above this value
the occupation of the ground (black line) and first excited (red line) state in Fig. \ref{lzt}(a)
is nearly constant. Variation of the occupation is found when the $\sin\nu t$ function
acquires values lower than $-0.8$. In Fig. \ref{lzt}(a) transitions from the ground to the excited
state after each sweep of the AC field accumulate which results in a constructive LZSM interference.
For contrast, in Fig. \ref{lzt}(b) we took the driving frequency between the two rightmost peaks of Fig. \ref{asy3}. Here the effects of the transitions do not accumulate but cancel at each sweep.}

 \begin{figure}
 \begin{tabular}{ccc}
 \includegraphics[height=.5\columnwidth]{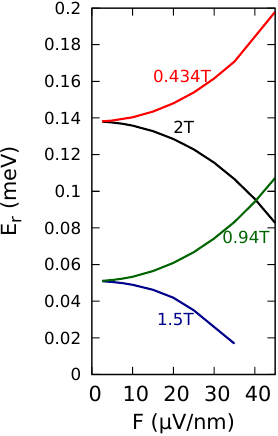}\put(-52,125){(a)} & 
 \includegraphics[height=.5\columnwidth]{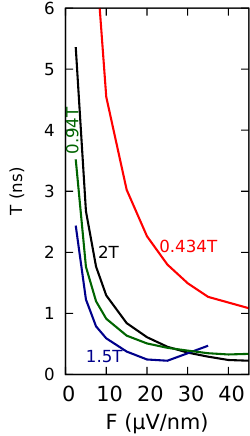}\put(-52,125){(b)} & \includegraphics[height=.49\columnwidth]{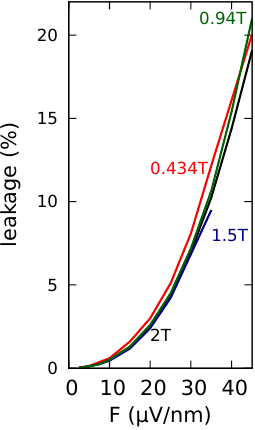} \put(-52,125){(c)}
 \end{tabular}
\caption{
The results for weakly asymmetric double quantum dot with $V_1=51$ meV 
and $s_x=29.64$ nm. 
(a) shows the position of the spin-flip singlet-triplet resonance in the AC
electric field as a function of its amplitude for selected
values of the external magnetic field. (b) The spin-flip time as a function of the amplitude in the first-order transition.  The line for $1.5$~T is interrupted since the first order peak can not longer be resolved at large $F$. (c) The maximal share of the states other than the two lowest-energy ones in the time evolution of the wave function for the system driven by the external AC field.}
 \label{sz}
\end{figure}

Figure \ref{sz}(a) shows the position of the first order transition on the energy scale $E_r=h\nu_r$ as a function of the AC amplitude. Note that, the magnetic fields considered in Figs. \ref{sz}(a) and Fig. \ref{asy2} were chosen to produce equal singlet-triplet energy splitting before and after the avoided crossings, hence the coalescent lines at the limit of zero amplitude. 
The frequency shifts before and after the singlet-triplet avoided crossing have opposite signs but similar magnitudes. 
The line for $1.5$~T is interrupted above 35~$\mu$V/nm, since the resonance line can be hardly recognized in the spectrum [cf. Fig. \ref{asy2}(c)].

A strong AC amplitude induces appearance of other states than the lowest-energy couple in the time evolution. The maximal share of the higher energy states as a function of the amplitude is given in Fig. \ref{sz}(c) as the leakage of the wave function outside the lowest-energy singlet and triplet. The dependence was
determined at the first-order resonant transition and as we can see the leakage on the amplitude is similar for all the magnetic fields considered. At the amplitude of $20~\mu$V/nm the maximal share of the higher energy states is about 2.5\%. 


The dependence of the spin-flip time on the amplitude for the first-order transition is plotted in Fig. \ref{sz}(b).
The minimal flip time of about 245 ps for $B=1.5$~T is obtained at the amplitude of 25~$\mu$V/nm. For even stronger amplitudes the flip time no longer decreases which is related to the potential sweeps across the singlet-triplet avoided crossing and the leakage of
the wave function to higher energy states for which the evolution no longer
have a two-state Rabi resonance character \cite{Sherman}. 

{\color{black}
Due to the very strong spin-orbital interactions in the present system the minimal EDSR-induced slip flip time of 245 ps is very short. Note that the fastest single-triplet transition  is obtained when the time evolution is not longer a two-level Rabi cycle with the presence of higher-excited states of about 5\% (see Fig. 13(c)).
However, even for the lowest considered AC field amplitude, when the spin transitions follow the Rabi cycle, the spin flip times of a few nanoseconds are still very low
 as compared to the values observed in quantum dots defined in GaAs or Si.
 The corresponding electron spin inversion time mediated via 
the spin-orbit coupling in electrostatic double quantum dots in GaAs is 106 ns \cite{Nowack2007}.
Faster spin-inversion times of about 50 ns have been achieved with the microwave chip generating magnetic field locally in the GaAs quantum dots \cite{Koppens2006} in an electron spin resonance setup. For GaAs quantum dots with the conduction band formed by s-type electron orbitals the spin inversion can also be achieved in the random field of nuclear spins \cite{Laird09} but the Rabi oscillations in the hyperfine field are very slow with the characteristic time of about 1 microsecond. The effects of the hyperfine interactions should
be negligible in the present system due to the d-type of the Ti orbitals that span the space for
the confined states. 
In silicon the hyperfine field can be minimized in isotopically enriched $^{28}$Si 
and the spin-flip time mediated by spin-valley coupling is of the order of 0.5 microsecond \cite{Klemt}. Advanced quantum registers of spin-qubits in silicon quantum dots are characterized
by faster spin-flip times of $\sim 100$ ns \cite{Philips}. These devices \cite{Philips} 
are equipped with integrated micromagnets that form a gradient of the magnetic field which mediates
the spin flips in external AC electric field that sweeps the confined electrons through the inhomogenous magnetic field region.}

\subsection{\color{black} Orientation of the fields and the interdot spacing}
\begin{figure}
\begin{tabular}{ll}
 \includegraphics[height=.6\columnwidth]{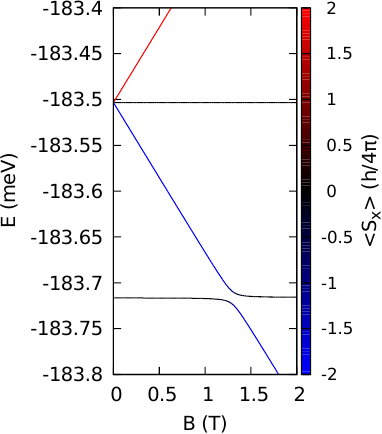}\put(-40,34){(a)}\end{tabular}
 \includegraphics[height=.5\columnwidth]{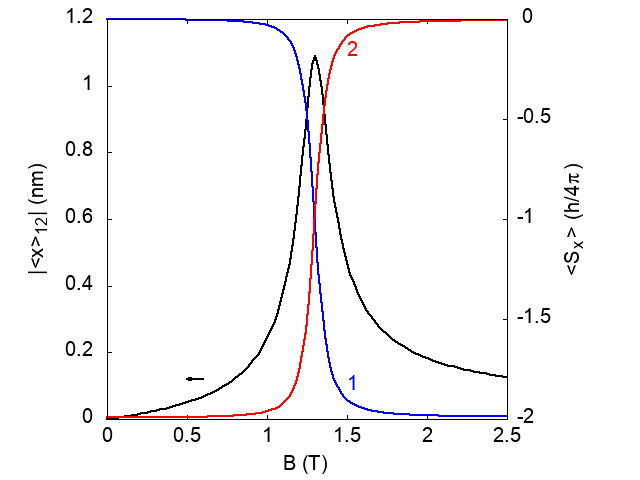}\put(-40,34){(b)} 
\caption{\color{black}
(a) Same as Fig. \ref{asy1} only with the magnetic field oriented in the $x$ direction $(B,0,0)$ and with 
the color on panel (b) showing the $S_x$ component of the total spin for the ground state (1) and the first excited state (2).
} 
 \label{asy1wbpx}
\end{figure}

\begin{figure}
 \begin{tabular}{l}
\includegraphics[height=.5\columnwidth]{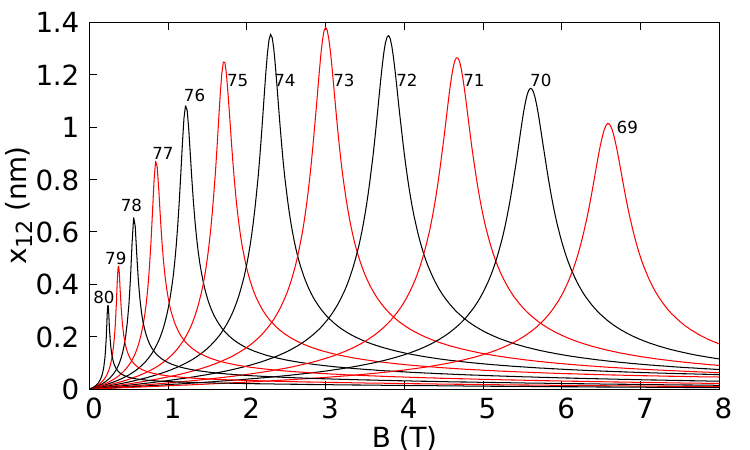} \put(-39,105){(a)} \\
\includegraphics[height=.5\columnwidth]{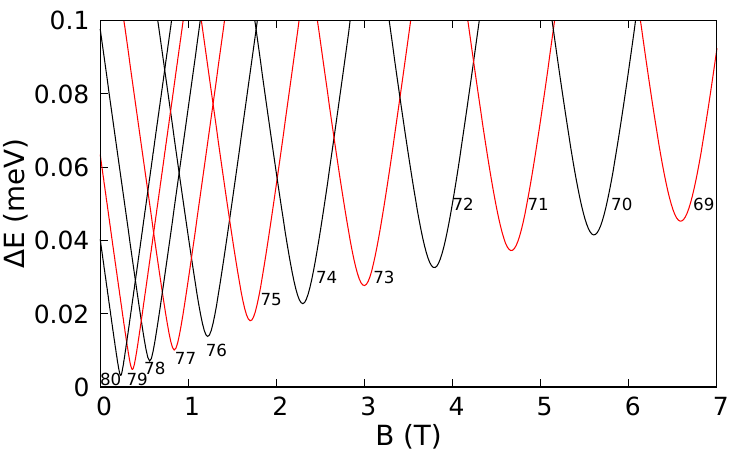} \put(-39,105){(b)} \\
\end{tabular} 
\caption{\color{black}
(a) Singlet-triplet matrix element as function of the perpendicular magnetic field (0,0,B) for $V_0=50$ meV and  $V_1=51$ meV and a range of values 
of the distance between the Gaussian well centers $s_x=n a$, where $a=0.39$nm is the lattice constant
and the value of $n$ is given in the figure. The plots in Fig. 1-14 correspond to $n=76$ 
while $n=80$ is considered in Figs. 1-3.
(b) The singlet-triplet energy splitting for same parameters as in (a).
The color of the lines was used to make the distinction of the data for separate $n$ easier.
}
 \label{wsx}
\end{figure}

{\color {black}
In the paper we assume that the magnetic field is oriented perpendicular
to the STO/LAO interface and that the AC electric field is oriented parallel
to the axis of the double dot system.
 For 
a choice of the orientation within the plane of confinement the orbital
effects of the magnetic field are absent. 
The energy spectrum for the weakly asymmetric system of Fig. \ref{asy1} 
and the magnetic field oriented parallel to the $x$ axis is displayed
in Fig. \ref{asy1wbpx}(a). One notices that for the in-plane magnetic field
orientation the curvature of the energy levels as a function of $B$ is missing
outside the singlet-triplet avoided-crossing. The avoided crossing is shifted
to higher values of the magnetic field by about 0.1T. The spin polarization
of the lowest-states as well as the dipole matrix element given in Fig. \ref{asy1wbpx}(b)
are very similar to the results in the perpendicular magnetic field (Fig. \ref{asy1}(b))
with the exception of a small shift of the $x_{12}$ maximum due to the shift on the $B$ scale of the avoided crossing.
The values of the matrix elements, the energy splittings and the $S_x$ spin components 
are given in Tables IV, V and VI, respectively. Overall, the orientation of the magnetic field turns out to be of a secondary importance.

The AC field oriented in the $x$ direction chosen in this paper produces large dipole matrix
elements due to the spatial extension of the charge system (see Fig. 1). The field oriented
parallel to the $y$ axis produces the matrix elements which are by two or three orders of magnitude 
smaller (Table VII). The present 2D model does not allow for a straightforward incorporation of the
AC electric field variation in the direction perpendicular to the plane of confinement.
However, as we argued in Ref. \cite{Szafran2023} the EDSR can be understood as a translation
of the electron motion induced by AC field to an effective magnetic field that is -- as in Rashba type coupling -- orthogonal to the direction of the induced electron spatial movement. 
The effective field needs to be orthogonal to the external magnetic field polarizing the 
electron spins. Therefore, the AC modulation of the electric potential perpendicular to the 
plane of confinement should lead to the transitions between the states polarized in this direction. We do not expect the effect to be stronger than for the field oriented along the axis of the dots
due to the very strong confinement of the wave functions in the $z$ direction (see Fig. 14 of Ref. \cite{Szafran2023}).

The effect of the distance between the Gaussian quantum well centers 
on the dipole matrix element and the singlet-triplet avoided crossing 
is given in Fig. \ref{wsx}(a,b). The curves in Fig. \ref{wsx} are denoted
by an integer $n$, where the spacing between the centers is taken as $s_x=na$ where $a = 0.39$ nm is the lattice
constant.
The values of $s_x=29.64$ nm and 31.2 nm used in the precedent section correspond
to $n=76$ and $n=78$, respectively. 
The interdot spacing has a pronounced effect on the position of the singlet-triplet
avoided crossing that produces the maximum of the dipole matrix element.
For a large interdot spacing the singlet and triplet tend to degeneracy that
occurs for strictly separated electrons. A small magnetic field 
promotes the spin-polarized triplet to the ground-state. For a small values of $n$ the systems
tends to a single elongated quantum dot and a large magnetic field is required
for a ground-state spin transition. The maximum of the dipole matrix element
is a non-monotonic function of the interdot spacing. In the limit of decoupled
quantum dots with absent interdot tunnel coupling the dipole matrix decreases.
On the other hand for a small value of $n$ the dipole matrix element cannot reach
large values due to a limited extension of the wave functions. The magnetic field
that is chosen for the spin manipulation needs to be placed off the avoided crossing
since the spin states are evenly mixed at the center of the avoided crossing.
Note, that the width of the peaks in Fig. \ref{wsx}(a) grows with decreasing $n$
which will affect the experimental workpoint. 
}

\begin{table}
\begin{tabular}{ccccc}
$n$& $|x_{1n}|$(0.434T)&$|x_{1n}|$(0.94T)&$|x_{1n}|$(1.5T)&$|x_{1n}|$(2T)\\ \hline
1&2.59&2.58&0.52&0.43\\
2&0.042&0.193&0.46&0.18\\
3&$3.5\times 10^{-4}$&$7.7\times 10^{-3}$&$8.2\times 10^{-4}$&$1.0\times 10^{-3}$\\
4&$2.1\times 10^{-2}$&$1.2\times 10^{-2}$&$1.0\times 10^{-2}$&$6.2\times 10^{-3}$\\
5&$8.05$&$8.02$&$1.77$&$0.69$\\
6&$3.49$&$3.48$&$0.79$&$0.33$\\
\end{tabular}
\caption{\color{black} The absolute value of the dipole matrix elements (in nanometers) for the six lowest energy states and the parameters
of Fig. \ref{asy1} at 4 selected values of the magnetic field $(B,0,0)$ that are used in Fig. 6.}
\end{table}

\begin{table}
\begin{tabular}{ccccc}
$n$& $\Delta E_{1n}$(0.434T)&$\Delta E_{1n}$(0.94T)&$\Delta E_{1n}$(1.5T)&$\Delta E_{1n}$(2T)\\ \hline
1&0&0&0&0\\
2& 0.142 & 0.0598& 0.0363 & 0.116\\
3& 0.213 & 0.214 & 0.0245& 0.329 \\ 
4& 0.284 & 0.367& 0.493& 0.656 \\
5& 1.318& 1.318& 1.352 & 1.433 \\
6& 2.640& 2.640& 2.674 & 2.755\\
\end{tabular}
\caption{\color{black} The energy spacing from the ground-state (in meV) for six lowest energy states and the parameters of Fig. \ref{asy1}  in the magnetic field $(B,0,0)$ -- cf. Table I for perpendicular magnetic field.}
\end{table}

\begin{table}
\begin{tabular}{ccccc}
$n$& $S_x$(0.434T)&$S_x$(0.94T)&$S_x$(1.5T)&$S_x$(2T)\\ \hline
1&$-2.3\times 10^{-3}$ & $-2.13\times 10^{-2}$ & -1.905 & -1.981 \\
2& -1.989& -1.970& $-8.63\times 10^{-2}$ &$1.06\times 10^{-2}$\\
3&  $-1.3\times 10^{-7}$ & $-8.26\times 10^{-7}$& $1.6\times 10^{-6} $&$2.3\times 10^{-6}$\\
4& 1.992&1.192&1.992&1.991\\
5&$3.8\times 10^{-4}$&$5.0\times 10^{-4}$&$6.6\times 10^{-4}$&$8.3\times 10^{-4}$ \\
6&$-1.5\times 10^{-4}$&$-1.8.0\times 10^{-4}$&$-2.0\times 10^{-4}$&$-2.3\times 10^{-4}$ \\
\end{tabular}
\caption{\color{black} The $S_x$ component of the spin in $\hbar/2$ units for the six lowest energy states and the parameters of Fig. \ref{asy1}  and magnetic fields $(B,0,0)$ -- cf. Table I for perpendicular magnetic field and $S_z$ component of the spin.}
\end{table}

\begin{table}
\begin{tabular}{cc}
$B$&$|y_{12}(B_z)|$\\ \hline
0.434T & $2.19\times 10^{-5}$\\
0.94T &$7.52\times 10^{-5}$\\
1.5T &$1.62\times 10^{-4}$\\
2T & $2.57\times 10^{-4}$\\
\end{tabular}
\caption{ \color{black}The singlet-triplet transition dipole matrix element in nanometers for the AC field
oriented in $y$ direction instead of the $x$ direction (the double dot system axis)
for the perpendicular magnetic field.}
\end{table}


\subsection{Optimization of the dipole matrix moment}

In the Rabi oscillations regime the rate of the first-order spin-flip transition is proportional to the value of the dipole matrix element. The latter is strictly zero for an ideally symmetric double quantum dots. In the precedent section we described the results obtained in a system with a weak asymmetry but the value of the matrix element can be optimized with the system parameters.
 \begin{figure}
 \begin{tabular}{llll}
\includegraphics[trim=0 0 30 0,clip,height=.5\columnwidth]{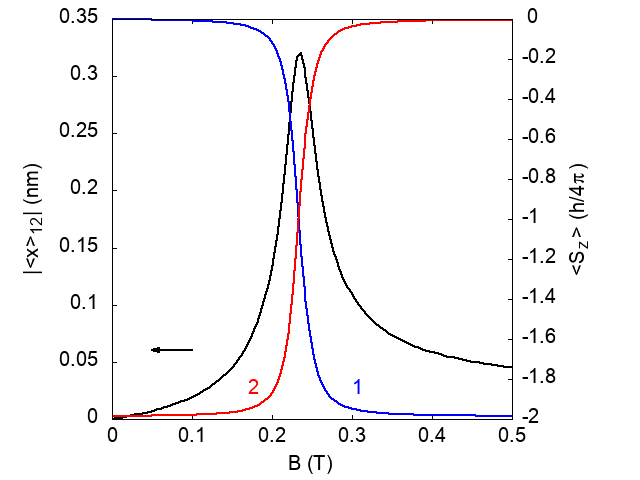} \put(-39,105){(a)} \\
 \includegraphics[trim=0 0 30 0,clip,height=.5\columnwidth]{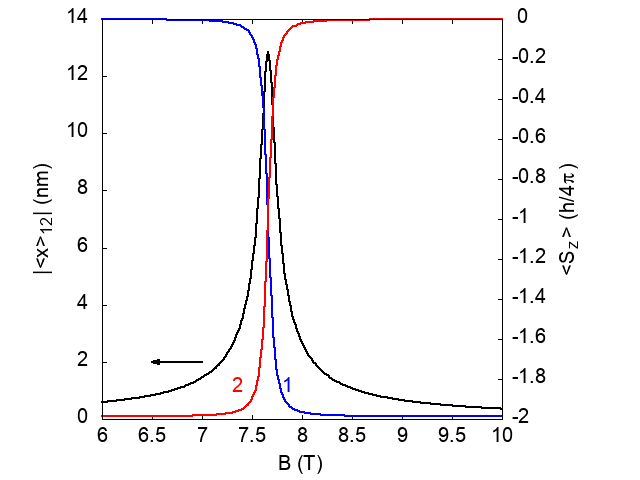} \put(-39,105){(b)}  \\
 \includegraphics[trim=0 0 30 0,clip,height=.5\columnwidth]{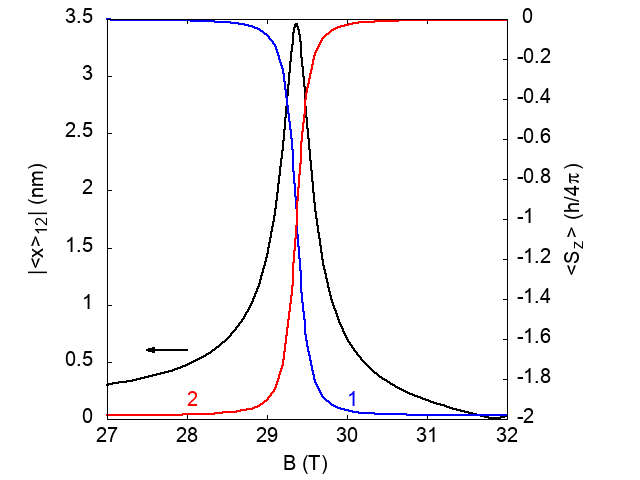} 
\put(-39,105){(c)} 
\end{tabular} 
\caption{
Same as Fig. \ref{asy1}(b) but for $s_x=31.2$ nm and $V_1=51$ meV (a), 
$V_1=54$ meV (b) and $60$ meV (c).}
 \label{ac80}
\end{figure}

The dipole matrix element for $s_x=29.64$~nm is displayed in Fig. \ref{asy1}(b).
In a system with larger interdot barrier ($s_x=31.2$~nm), but still with the small asymmetry $V_1=51$~meV as in Fig. \ref{asy1}(b), the singlet-triplet avoided crossing occurs at lower magnetic field of $B=0.23$~T 
[cf. Fig. \ref{ac80}(a)] than for the stronger interdot coupling due to a reduction of the exchange energy [Fig. \ref{v1}(a)]. 
The width of the singlet-triplet avoided crossing is 3.2~$\mu$eV for $s_x=31.2$~nm. 
The maximal value of the dipole matrix element in the center of the avoided crossing
is about $0.323$~nm, e.g. about 3 times lower than for $s_x=29.64$~nm [Fig. \ref{asy1}(b)].

The position of the singlet-triplet avoided crossing and the maximal value of the dipole matrix element strongly depends on the asymmetry of the confinement potential. 
In Fig. \ref{ac80}(b) and (c) we plotted the average spins and the dipole matrix element near the singlet-triplet avoided-crossing for $s_x=31.2$ nm with $V_1=54$ meV (b) and $V_1=60$ meV (c). For $V_1=54$ meV [Fig.~\ref{ac80}(b)] the maximal value of the dipole matrix element is 40 times larger than for the small asymmetry $V_1=51$~meV  [Fig.~\ref{ac80}(a)] and about 4 times larger than for the large asymmetry $V_1=60$~meV [Fig.\ref{ac80}(c)].

\begin{figure}
\begin{tabular}{ll}
  \includegraphics[trim=0 0 0 0,clip,height=.55\columnwidth]{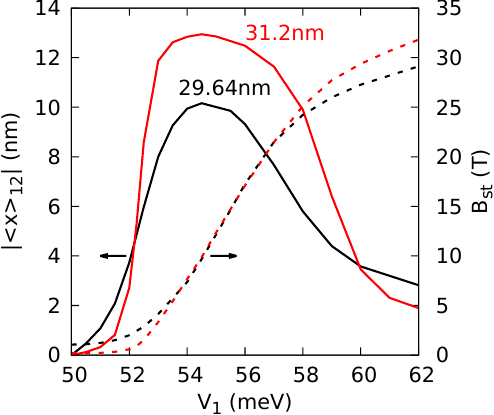}
  \put(-135,120){(a)}\\
  \includegraphics[trim=0 0 0 0,clip,height=.55\columnwidth]{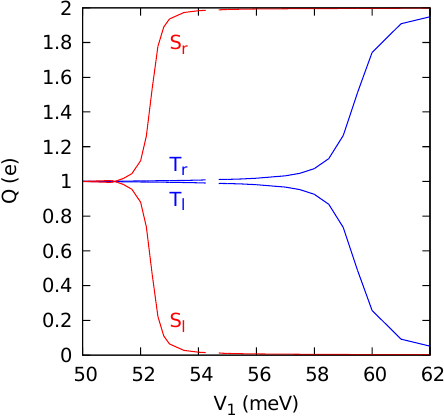}
  \put(-115,120){(b)}\\
  \end{tabular}
  \caption{
  (a) The solid lines show the dipole matrix element for the transition between the two lowest-energy states at the center of the singlet-triplet avoided crossing.
  The dashed lines show the position of the avoided crossing (right vertical scale).
  The red and black lines correspond to the $s_x$ parameter equal to $31.2$~nm and $29.64$~nm, respectively. 
  (b) The charge localized in the right and left quantum dots for 
  $s_x=31.2$ nm in the singlet (red lines) and triplet (blue lines) 
  for $B=10$~T. The gap in the lines was left for the avoided crossing 
  of the singlet and triplet energy levels for the clarity of the plot.
  The ground-state is singlet (triplet) at the right(left)-hand side of the figure. }
  \label{wv1}
\end{figure}

The solid lines in Fig. \ref{wv1}(a) show the absolute value of the dipole element
at the center of the singlet-triplet avoided crossing as a function of the right-dot potential $V_1$.
The asymmetry of the confinement potential pushes the avoided crossing to higher values of the magnetic field (dashed lines referred to the right axis). The dependence of the matrix element on the asymmetry is  non-monotonic which is related to the charge distribution in the singlet  and triplet states. Fig. \ref{wv1}(b) shows the charge in the left and right quantum dot
calculated for $B=10$~T -- near the value of the magnetic field for which the maximum 
of the dipole matrix element is achieved in Fig. \ref{wv1}(a) for $V_1=54$ meV.
The gap in the lines of Fig. \ref{wv1}(b) corresponds to the region of the singlet-triplet avoided crossing as a function of $V_1$ that we skipped for the clarity of the plot.
At the left (right) hand side of the avoided crossing the ground-state for $B=10$~T is the spin triplet (singlet).

Comparing Fig. \ref{wv1}(a) and Fig. \ref{wv1}(b) we can conclude that
the maximal value of the dipole matrix element is achieved for $V_1$ which 
on the one hand localizes the singlet entirely in the deeper dot but for which
the triplet charge distribution is still evenly distributed in both dots. 
In the region of the avoided crossing the two-lowest energy states
have hybridized wave functions that due to the contribution of the triplet are delocalized over both dots which produces a large value of the dipole matrix element. For large $V_1$, on the other hand, both singlet and triplet states are confined within a single deeper quantum dot with potential perturbed by the potential of the left Gaussian. Due to this perturbation the 
dipole matrix element is non-zero. Since the system is localized in a single dot the matrix element does not achieve values as large as for the delocalized triplet state.
Comparing Fig.~\ref{v1}(b) and Fig.~\ref{wv1}(a) we also notice that
at higher $B$ the triplet state is more easily localized in the deeper
dot: for $V_1=60$meV the charge in the right dot is $1.2e$ at B=0,
but already $1.7e$ at $B=10$~T.
\begin{figure}
  \includegraphics[trim=0 0 0 0,clip,height=.6\columnwidth]{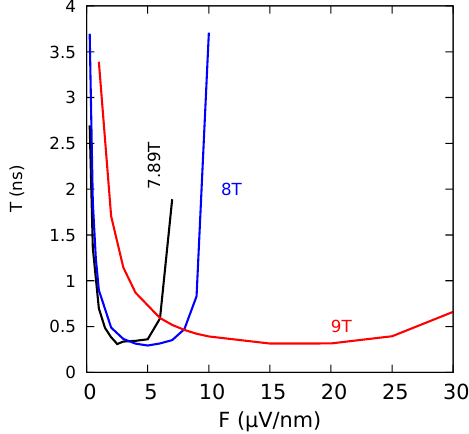}
  \caption{Singlet-triplet transition time for the double-dot system with $s_x=31.2$~nm and $V_1=54$~meV driven by the AC field of $eFx\sin(\nu t)$ for three values of the magnetic field corresponding to different values of the dipole moment, see Fig. \ref{ac80}(b).}
  \label{cz80}
\end{figure}

For the studies of the transition times we chose $V_1=54$~meV -- near the maximum
of the dipole matrix element [Fig. \ref{wv1}(a)] and three values of the magnetic field after the singlet-triplet avoided crossing, where the ground-state is the spin triplet -- $7.9$~T, $8$~T and $9$~T -- see Fig.~\ref{ac80}(b). 
The singlet-triplet transition times as functions of the AC amplitude is given 
in Fig. \ref{cz80}. The transition times reach minima for lower values of the
AC field amplitude than in Fig. \ref{sz}(b) and start to grow for the $F$ values that
exceeds the one for the LZSM transition regime. 
The LZSM regime is open for the amplitudes of $F=1.95~\mu$V, 2.7$~\mu$V and 9.5~$\mu$V/nm for $B=7.89$~T, $8$~T and $9$~T, respectively.
Note, that for each value of the magnetic field the minimum spin flip time of about 330~ps is achieved for some value of the AC field amplitude. Since the amplitudes of the AC field near the minimal spin flip time are smaller than in Fig. \ref{sz}(b), so are the wave function leakage values near the minimum which are only 0.5\%, 0.02\%, 0.0043\% for $F=17.5~\mu$V/nm, $5~\mu$V/nm, and $2.43~\mu$V/nm,
corresponding to $9$~T and $8$~T, and $B=7.89$~T, respectively.

\subsection{Spin flips for an ideally symmetric system}

\begin{figure}
\begin{tabular}{l}
 \includegraphics[width=.96\columnwidth]{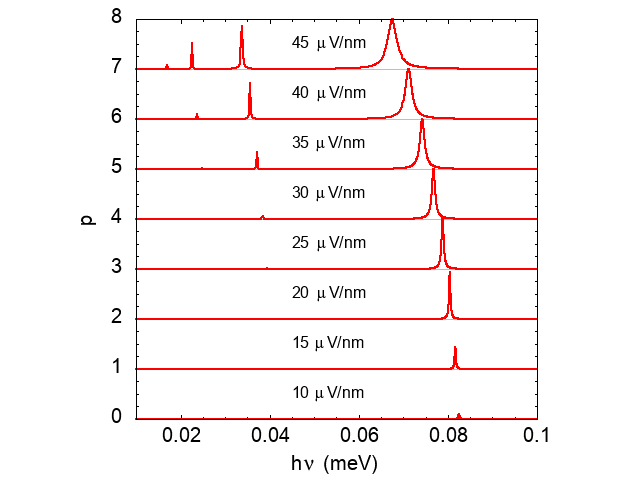} 
  \end{tabular}
\caption{Same as Fig. \ref{asy2} only for a symmetric system with $V_0=V_1=50$~meV
at $B=2$~T.}
 \label{spar}
\end{figure}

For completeness we present the results for an ideally symmetric system,
for which the direct spin-flip transition is forbidden by the selection rule. 
Fig. \ref{spar} shows the probability of the spin-flip-singlet for $s_x=29.64$~nm and the external magnetic field set at $2$~T, for a simulation
lasting $5$~ns and starting from the ground-state triplet as the initial state. 
The subsequent lines at the upper part of  Fig. \ref{spar} correspond
to a growing amplitude of the AC field. The most pronounced peak that is observed
in Fig. \ref{spar} at weak AC field amplitude occurs at roughly half the singlet-triplet energy splitting that at the field of $2$~T is equal to $\Delta E_{ST}=0.1663$~meV [cf. Fig. \ref{spesy}(b)]. 
The spin-flip occurs for AC frequency set at this peak by a two-photon second-order process. Also, no third-order (three-photon) peak at 1/3 of $\Delta E_{ST}$ is observed.  For higher amplitude the position of this peak gets red-shifted and 
the peaks corresponding to even denominators, e.g. 1/4 and 1/6 of $\Delta E_{ST}$ are formed.

\begin{figure}
\begin{tabular}{lll}
 \includegraphics[width=.43\columnwidth]{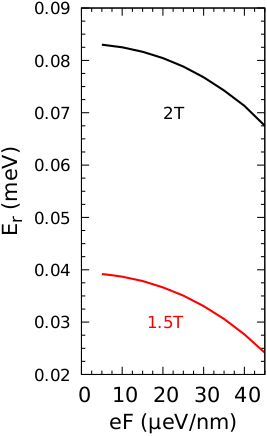} \put(-15,160){(a)}  &
 \includegraphics[width=.40\columnwidth]{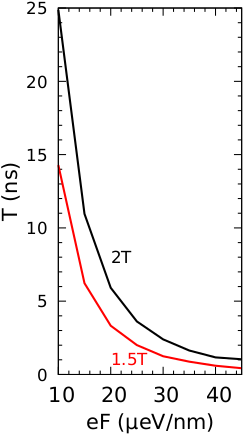} \put(-15,160){(b)}  
  \end{tabular}
\caption{
(a) The position of the second-order (two-photon) singlet-triplet resonance
in the AC field of $eFx\sin(\nu t)$ for two values of the magnetic field.
For both $B$ values the ground-state of the system is the spin triplet.
(b) The spin-flip transition time at the singlet-triplet resonance as a function
of the driving field amplitude. Results for a symmetric double quantum dots with $s_x=29.64$~nm and $V_0=V_1=50$~meV.}
 \label{symed}
\end{figure}

The study of the AC driven spin-flips in a symmetric system with $s_x=29.64$~nm is summarized in Fig. \ref{symed} for two values of the external magnetic field $B=1.5$~T and $B=2$~T, both corresponding to the ground-state triplet.
 The position of the resonant second-order transition on the
energy scale is redshifted with the amplitude [Fig. \ref{symed}(a)]. 
In the studied range of the AC amplitudes -- the spin-flip time is a decreasing
function of the amplitude [Fig. \ref{symed}(b)].  
For $1.5$~T the LZSM regime opens at 45~$\mu$eV/nm, and for $2$~T at still higher amplitudes which are outside the range studies in Fig. \ref{symed}.

\section{Summary and conclusions}
We have studied two electrons confined in a lateral double quantum dot defined within the two-dimensional electron gas on the (001)-oriented LAO/STO surface.
Utilizing the three-band model of $3d$-electrons localized at Ti ions on a square lattice, we have analyzed the energy spectrum of the system, paying particular attention to the singlet-triplet avoided crossing induced by the SO interaction. The calculated eigenstates have been employed as initial conditions for time-dependent calculations. In the magnetic field range close to the avoided crossing, the spin-flip singlet-triplet transition induced by an AC electric field has been analyzed in details in the context of the electric dipole spin resonance experiment.

We have demonstrated that for a fully symmetric system the first order singlet-triplet transition are forbidden due to the parity symmetry of the Hamiltonian which results in zeroing of the dipole matrix elements. To induce the spin-flip transition we introduce the asymmetry in the confinement potential in the form of unequal depths of the Gaussian quantum wells. Our simulations for a weakly asymmetric system show that the spin-flip singlet-triplet transition has the character of the Rabi oscillations for a low AC field amplitude, with minimal flip time of about 245 ps. Interestingly, when we increase the AC filed amplitude the frequency of the transition is blueshifted (redshifted) for the magnetic field below (above) the single-triplet avoided crossing. Moreover, we have noticed that for a magnetic field above the anticrossing the regular transition spectrum is highly deteriorated at a low frequency regime and the high AC field amplitude. This behavior has been explained as resulting from the Landau-Zener-Stueckelberg-Majorana transitions between the ground and first excited states which occur in the system when the electric field is large enough to drive the system across the avoided crossing.

The optimization of system geometrical parameters shows that the appropriate  potential asymmetry between the dots highly reduces the leakage of transition to the higher states which can reach the value of about 0.05\% with the spin flip time at the order of $330$~ps.
Finally our simulations for a fully symmetric system provide insight into the possibility of the spin-flip transition by the multiphoton processes. The singlet-triplet transition studied here will be relevant for future EDSR experiments aimed at reading the spin state of the dot and paving the way to a spin-orbit qubit at the oxides interface.

\section{ACKNOWLEDGEMENT}
This work is financed by the Horizon Europe EIC Pathfinder under the grant IQARO number 101115190 titled "Spin-orbitronic quantum bits in reconfigurable 2D-oxides"
and partly supported by program „Excellence initiative – research university” for the AGH University. 
Computing infrastructure PLGrid (HPC Centers: ACK Cyfronet AGH) within computational grant no. PLG/2024/017175 was used.

\bibliography{ref.bib}

\begin{thebibliography}{48}%
\makeatletter
\providecommand \@ifxundefined [1]{%
 \@ifx{#1\undefined}
}%
\providecommand \@ifnum [1]{%
 \ifnum #1\expandafter \@firstoftwo
 \else \expandafter \@secondoftwo
 \fi
}%
\providecommand \@ifx [1]{%
 \ifx #1\expandafter \@firstoftwo
 \else \expandafter \@secondoftwo
 \fi
}%
\providecommand \natexlab [1]{#1}%
\providecommand \enquote  [1]{``#1''}%
\providecommand \bibnamefont  [1]{#1}%
\providecommand \bibfnamefont [1]{#1}%
\providecommand \citenamefont [1]{#1}%
\providecommand \href@noop [0]{\@secondoftwo}%
\providecommand \href [0]{\begingroup \@sanitize@url \@href}%
\providecommand \@href[1]{\@@startlink{#1}\@@href}%
\providecommand \@@href[1]{\endgroup#1\@@endlink}%
\providecommand \@sanitize@url [0]{\catcode `\\12\catcode `\$12\catcode
  `\&12\catcode `\#12\catcode `\^12\catcode `\_12\catcode `\%12\relax}%
\providecommand \@@startlink[1]{}%
\providecommand \@@endlink[0]{}%
\providecommand \url  [0]{\begingroup\@sanitize@url \@url }%
\providecommand \@url [1]{\endgroup\@href {#1}{\urlprefix }}%
\providecommand \urlprefix  [0]{URL }%
\providecommand \Eprint [0]{\href }%
\providecommand \doibase [0]{http://dx.doi.org/}%
\providecommand \selectlanguage [0]{\@gobble}%
\providecommand \bibinfo  [0]{\@secondoftwo}%
\providecommand \bibfield  [0]{\@secondoftwo}%
\providecommand \translation [1]{[#1]}%
\providecommand \BibitemOpen [0]{}%
\providecommand \bibitemStop [0]{}%
\providecommand \bibitemNoStop [0]{.\EOS\space}%
\providecommand \EOS [0]{\spacefactor3000\relax}%
\providecommand \BibitemShut  [1]{\csname bibitem#1\endcsname}%
\let\auto@bib@innerbib\@empty
\bibitem [{\citenamefont {Burkard}\ \emph {et~al.}(2023)\citenamefont
  {Burkard}, \citenamefont {Ladd}, \citenamefont {Pan}, \citenamefont
  {Nichol},\ and\ \citenamefont {Petta}}]{Burkard2023}%
  \BibitemOpen
  \bibfield  {author} {\bibinfo {author} {\bibfnamefont {G.}~\bibnamefont
  {Burkard}}, \bibinfo {author} {\bibfnamefont {T.~D.}\ \bibnamefont {Ladd}},
  \bibinfo {author} {\bibfnamefont {A.}~\bibnamefont {Pan}}, \bibinfo {author}
  {\bibfnamefont {J.~M.}\ \bibnamefont {Nichol}}, \ and\ \bibinfo {author}
  {\bibfnamefont {J.~R.}\ \bibnamefont {Petta}},\ }\bibfield  {title} {\enquote
  {\bibinfo {title} {Semiconductor spin qubits},}\ }\href {\doibase
  10.1103/RevModPhys.95.025003} {\bibfield  {journal} {\bibinfo  {journal}
  {Rev. Mod. Phys.}\ }\textbf {\bibinfo {volume} {95}},\ \bibinfo {pages}
  {025003} (\bibinfo {year} {2023})}\BibitemShut {NoStop}%
\bibitem [{\citenamefont {Nadj-Perge}\ \emph {et~al.}(2010)\citenamefont
  {Nadj-Perge}, \citenamefont {Frolov}, \citenamefont {Bakkers},\ and\
  \citenamefont {Kouwenhoven}}]{Nadj-Perge2010}%
  \BibitemOpen
  \bibfield  {author} {\bibinfo {author} {\bibfnamefont {S.}~\bibnamefont
  {Nadj-Perge}}, \bibinfo {author} {\bibfnamefont {S.~M.}\ \bibnamefont
  {Frolov}}, \bibinfo {author} {\bibfnamefont {E.~P. A.~M.}\ \bibnamefont
  {Bakkers}}, \ and\ \bibinfo {author} {\bibfnamefont {L.~P.}\ \bibnamefont
  {Kouwenhoven}},\ }\bibfield  {title} {\enquote {\bibinfo {title} {Spin-orbit
  qubit in a semiconductor nanowire},}\ }\href@noop {} {\bibfield  {journal}
  {\bibinfo  {journal} {Nature}\ }\textbf {\bibinfo {volume} {468}},\ \bibinfo
  {pages} {1084} (\bibinfo {year} {2010})}\BibitemShut {NoStop}%
\bibitem [{\citenamefont {Giglberger}\ \emph {et~al.}(2007)\citenamefont
  {Giglberger}, \citenamefont {Golub}, \citenamefont {Bel'kov}, \citenamefont
  {Danilov}, \citenamefont {Schuh}, \citenamefont {Gerl}, \citenamefont
  {Rohlfing}, \citenamefont {Stahl}, \citenamefont {Wegscheider}, \citenamefont
  {Weiss}, \citenamefont {Prettl},\ and\ \citenamefont {Ganichev}}]{Rashba}%
  \BibitemOpen
  \bibfield  {author} {\bibinfo {author} {\bibfnamefont {S.}~\bibnamefont
  {Giglberger}}, \bibinfo {author} {\bibfnamefont {L.~E.}\ \bibnamefont
  {Golub}}, \bibinfo {author} {\bibfnamefont {V.~V.}\ \bibnamefont {Bel'kov}},
  \bibinfo {author} {\bibfnamefont {S.~N.}\ \bibnamefont {Danilov}}, \bibinfo
  {author} {\bibfnamefont {D.}~\bibnamefont {Schuh}}, \bibinfo {author}
  {\bibfnamefont {C.}~\bibnamefont {Gerl}}, \bibinfo {author} {\bibfnamefont
  {F.}~\bibnamefont {Rohlfing}}, \bibinfo {author} {\bibfnamefont
  {J.}~\bibnamefont {Stahl}}, \bibinfo {author} {\bibfnamefont
  {W.}~\bibnamefont {Wegscheider}}, \bibinfo {author} {\bibfnamefont
  {D.}~\bibnamefont {Weiss}}, \bibinfo {author} {\bibfnamefont
  {W.}~\bibnamefont {Prettl}}, \ and\ \bibinfo {author} {\bibfnamefont {S.~D.}\
  \bibnamefont {Ganichev}},\ }\bibfield  {title} {\enquote {\bibinfo {title}
  {Rashba and dresselhaus spin splittings in semiconductor quantum wells
  measured by spin photocurrents},}\ }\href {\doibase
  10.1103/PhysRevB.75.035327} {\bibfield  {journal} {\bibinfo  {journal} {Phys.
  Rev. B}\ }\textbf {\bibinfo {volume} {75}},\ \bibinfo {pages} {035327}
  (\bibinfo {year} {2007})}\BibitemShut {NoStop}%
\bibitem [{\citenamefont {Rashba}\ and\ \citenamefont
  {Efros}(2003)}]{Rashba2003}%
  \BibitemOpen
  \bibfield  {author} {\bibinfo {author} {\bibfnamefont {E.~I.}\ \bibnamefont
  {Rashba}}\ and\ \bibinfo {author} {\bibfnamefont {Al.~L.}\ \bibnamefont
  {Efros}},\ }\bibfield  {title} {\enquote {\bibinfo {title} {Orbital
  mechanisms of electron-spin manipulation by an electric field},}\ }\href
  {\doibase 10.1103/PhysRevLett.91.126405} {\bibfield  {journal} {\bibinfo
  {journal} {Phys. Rev. Lett.}\ }\textbf {\bibinfo {volume} {91}},\ \bibinfo
  {pages} {126405} (\bibinfo {year} {2003})}\BibitemShut {NoStop}%
\bibitem [{\citenamefont {Bychkov}\ and\ \citenamefont
  {Rashba}(1984)}]{Bychkov1984}%
  \BibitemOpen
  \bibfield  {author} {\bibinfo {author} {\bibfnamefont {Y.~A.}\ \bibnamefont
  {Bychkov}}\ and\ \bibinfo {author} {\bibfnamefont {E.~I.}\ \bibnamefont
  {Rashba}},\ }\bibfield  {title} {\enquote {\bibinfo {title} {Oscillatory
  effects and the magnetic susceptibility of carriers in inversion layers},}\
  }\href@noop {} {\bibfield  {journal} {\bibinfo  {journal} {J. Phys. C}\
  }\textbf {\bibinfo {volume} {17}},\ \bibinfo {pages} {6039} (\bibinfo {year}
  {1984})}\BibitemShut {NoStop}%
\bibitem [{\citenamefont {Khaetskii}\ \emph {et~al.}(2002)\citenamefont
  {Khaetskii}, \citenamefont {Loss},\ and\ \citenamefont
  {Glazman}}]{Khaetskii2002}%
  \BibitemOpen
  \bibfield  {author} {\bibinfo {author} {\bibfnamefont {A.~V.}\ \bibnamefont
  {Khaetskii}}, \bibinfo {author} {\bibfnamefont {D.}~\bibnamefont {Loss}}, \
  and\ \bibinfo {author} {\bibfnamefont {L.}~\bibnamefont {Glazman}},\
  }\bibfield  {title} {\enquote {\bibinfo {title} {Electron spin decoherence in
  quantum dots due to interaction with nuclei},}\ }\href {\doibase
  10.1103/PhysRevLett.88.186802} {\bibfield  {journal} {\bibinfo  {journal}
  {Phys. Rev. Lett.}\ }\textbf {\bibinfo {volume} {88}},\ \bibinfo {pages}
  {186802} (\bibinfo {year} {2002})}\BibitemShut {NoStop}%
\bibitem [{\citenamefont {Loss}\ and\ \citenamefont
  {DiVincenzo}(1998)}]{DiVincenzo1998}%
  \BibitemOpen
  \bibfield  {author} {\bibinfo {author} {\bibfnamefont {D.}~\bibnamefont
  {Loss}}\ and\ \bibinfo {author} {\bibfnamefont {D.~P.}\ \bibnamefont
  {DiVincenzo}},\ }\bibfield  {title} {\enquote {\bibinfo {title} {Quantum
  computation with quantum dots},}\ }\href {\doibase 10.1103/PhysRevA.57.120}
  {\bibfield  {journal} {\bibinfo  {journal} {Phys. Rev. A}\ }\textbf {\bibinfo
  {volume} {57}},\ \bibinfo {pages} {120--126} (\bibinfo {year}
  {1998})}\BibitemShut {NoStop}%
\bibitem [{\citenamefont {Mills}\ \emph {et~al.}(2022)\citenamefont {Mills},
  \citenamefont {Guinn}, \citenamefont {Gullans}, \citenamefont {Sigilitto},
  \citenamefont {Feldman}, \citenamefont {Nielsen},\ and\ \citenamefont
  {Petta}}]{Petta2022}%
  \BibitemOpen
  \bibfield  {author} {\bibinfo {author} {\bibfnamefont {A.~R.}\ \bibnamefont
  {Mills}}, \bibinfo {author} {\bibfnamefont {C.~R.}\ \bibnamefont {Guinn}},
  \bibinfo {author} {\bibfnamefont {M.~J.}\ \bibnamefont {Gullans}}, \bibinfo
  {author} {\bibfnamefont {A.~J.}\ \bibnamefont {Sigilitto}}, \bibinfo {author}
  {\bibfnamefont {M.~M.}\ \bibnamefont {Feldman}}, \bibinfo {author}
  {\bibfnamefont {E.}~\bibnamefont {Nielsen}}, \ and\ \bibinfo {author}
  {\bibfnamefont {J.~R.}\ \bibnamefont {Petta}},\ }\bibfield  {title} {\enquote
  {\bibinfo {title} {Two-qubit silicon quantum processor with operation
  fidelity exceeding 99 \%},}\ }\href {\doibase 10.1103/PhysRevB.75.035327}
  {\bibfield  {journal} {\bibinfo  {journal} {Sci. Adv.}\ }\textbf {\bibinfo
  {volume} {8}},\ \bibinfo {pages} {5130} (\bibinfo {year} {2022})}\BibitemShut
  {NoStop}%
\bibitem [{\citenamefont {Joshua}\ \emph {et~al.}(2012)\citenamefont {Joshua},
  \citenamefont {Pecker}, \citenamefont {Ruhman}, \citenamefont {Altman},\ and\
  \citenamefont {Ilani}}]{Joshua2012}%
  \BibitemOpen
  \bibfield  {author} {\bibinfo {author} {\bibfnamefont {A.}~\bibnamefont
  {Joshua}}, \bibinfo {author} {\bibfnamefont {S.}~\bibnamefont {Pecker}},
  \bibinfo {author} {\bibfnamefont {J.}~\bibnamefont {Ruhman}}, \bibinfo
  {author} {\bibfnamefont {E.}~\bibnamefont {Altman}}, \ and\ \bibinfo {author}
  {\bibfnamefont {S.}~\bibnamefont {Ilani}},\ }\bibfield  {title} {\enquote
  {\bibinfo {title} {A universal critical density underlying the physics of
  electrons at the $\mathrm{LaAlO}{}_{3}/\mathrm{SrTiO}{}_{3}$ interface},}\
  }\href@noop {} {\bibfield  {journal} {\bibinfo  {journal} {Nat. Commun.}\
  }\textbf {\bibinfo {volume} {3}},\ \bibinfo {pages} {1129} (\bibinfo {year}
  {2012})}\BibitemShut {NoStop}%
\bibitem [{\citenamefont {Maniv}\ \emph {et~al.}(2015)\citenamefont {Maniv},
  \citenamefont {Ben~Shalom}, \citenamefont {Mograbi}, \citenamefont
  {Pavelski}, \citenamefont {Goldstein},\ and\ \citenamefont
  {Dagan}}]{Maniv2015}%
  \BibitemOpen
  \bibfield  {author} {\bibinfo {author} {\bibfnamefont {E.}~\bibnamefont
  {Maniv}}, \bibinfo {author} {\bibfnamefont {M.}~\bibnamefont {Ben~Shalom}},
  \bibinfo {author} {\bibfnamefont {M.}~\bibnamefont {Mograbi}}, \bibinfo
  {author} {\bibfnamefont {A.}~\bibnamefont {Pavelski}}, \bibinfo {author}
  {\bibfnamefont {M.}~\bibnamefont {Goldstein}}, \ and\ \bibinfo {author}
  {\bibfnamefont {Y.}~\bibnamefont {Dagan}},\ }\bibfield  {title} {\enquote
  {\bibinfo {title} {Strong correlations elucidate the electronic structure and
  phase diagram of $\mathrm{LaAlO}{}_{3}/\mathrm{SrTiO}{}_{3}$ interface},}\
  }\href@noop {} {\bibfield  {journal} {\bibinfo  {journal} {Nat. Commun.}\
  }\textbf {\bibinfo {volume} {2}},\ \bibinfo {pages} {8239} (\bibinfo {year}
  {2015})}\BibitemShut {NoStop}%
\bibitem [{\citenamefont {Monteiro}\ \emph {et~al.}(2019)\citenamefont
  {Monteiro}, \citenamefont {Vivek}, \citenamefont {Groenendijk}, \citenamefont
  {Bruneel}, \citenamefont {Leermakers}, \citenamefont {Zeitler}, \citenamefont
  {Gabay},\ and\ \citenamefont {Caviglia}}]{Monteiro2019}%
  \BibitemOpen
  \bibfield  {author} {\bibinfo {author} {\bibfnamefont {A.~M. R. V.~L.}\
  \bibnamefont {Monteiro}}, \bibinfo {author} {\bibfnamefont {M.}~\bibnamefont
  {Vivek}}, \bibinfo {author} {\bibfnamefont {D.~J.}\ \bibnamefont
  {Groenendijk}}, \bibinfo {author} {\bibfnamefont {P.}~\bibnamefont
  {Bruneel}}, \bibinfo {author} {\bibfnamefont {I.}~\bibnamefont {Leermakers}},
  \bibinfo {author} {\bibfnamefont {U.}~\bibnamefont {Zeitler}}, \bibinfo
  {author} {\bibfnamefont {M.}~\bibnamefont {Gabay}}, \ and\ \bibinfo {author}
  {\bibfnamefont {A.~D.}\ \bibnamefont {Caviglia}},\ }\bibfield  {title}
  {\enquote {\bibinfo {title} {Band inversion driven by electronic correlations
  at the (111) $\mathrm{LaAlO}{}_{3}/\mathrm{SrTiO}{}_{3}$ interface},}\ }\href
  {\doibase 10.1103/PhysRevB.99.201102} {\bibfield  {journal} {\bibinfo
  {journal} {Phys. Rev. B}\ }\textbf {\bibinfo {volume} {99}},\ \bibinfo
  {pages} {201102} (\bibinfo {year} {2019})}\BibitemShut {NoStop}%
\bibitem [{\citenamefont {Pavlenko}\ \emph {et~al.}(2012)\citenamefont
  {Pavlenko}, \citenamefont {Kopp}, \citenamefont {Tsymbal}, \citenamefont
  {Sawatzky},\ and\ \citenamefont {Mannhart}}]{Pavlenko2012}%
  \BibitemOpen
  \bibfield  {author} {\bibinfo {author} {\bibfnamefont {N.}~\bibnamefont
  {Pavlenko}}, \bibinfo {author} {\bibfnamefont {T.}~\bibnamefont {Kopp}},
  \bibinfo {author} {\bibfnamefont {E.~Y.}\ \bibnamefont {Tsymbal}}, \bibinfo
  {author} {\bibfnamefont {G.~A.}\ \bibnamefont {Sawatzky}}, \ and\ \bibinfo
  {author} {\bibfnamefont {J.}~\bibnamefont {Mannhart}},\ }\bibfield  {title}
  {\enquote {\bibinfo {title} {Magnetic and superconducting phases at the
  $\mathrm{LaAlO}{}_{3}/\mathrm{SrTiO}{}_{3}$ interface: The role of
  interfacial ti 3$d$ electrons},}\ }\href {\doibase
  10.1103/PhysRevB.85.020407} {\bibfield  {journal} {\bibinfo  {journal} {Phys.
  Rev. B}\ }\textbf {\bibinfo {volume} {85}},\ \bibinfo {pages} {020407}
  (\bibinfo {year} {2012})}\BibitemShut {NoStop}%
\bibitem [{\citenamefont {Biscaras}\ \emph {et~al.}(2012)\citenamefont
  {Biscaras}, \citenamefont {Bergeal}, \citenamefont {Hurand}, \citenamefont
  {Grosset\^ete}, \citenamefont {Rastogi}, \citenamefont {Budhani},
  \citenamefont {LeBoeuf}, \citenamefont {Proust},\ and\ \citenamefont
  {Lesueur}}]{Biscaras2012}%
  \BibitemOpen
  \bibfield  {author} {\bibinfo {author} {\bibfnamefont {J.}~\bibnamefont
  {Biscaras}}, \bibinfo {author} {\bibfnamefont {N.}~\bibnamefont {Bergeal}},
  \bibinfo {author} {\bibfnamefont {S.}~\bibnamefont {Hurand}}, \bibinfo
  {author} {\bibfnamefont {C.}~\bibnamefont {Grosset\^ete}}, \bibinfo {author}
  {\bibfnamefont {A.}~\bibnamefont {Rastogi}}, \bibinfo {author} {\bibfnamefont
  {R.~C.}\ \bibnamefont {Budhani}}, \bibinfo {author} {\bibfnamefont
  {D.}~\bibnamefont {LeBoeuf}}, \bibinfo {author} {\bibfnamefont
  {C.}~\bibnamefont {Proust}}, \ and\ \bibinfo {author} {\bibfnamefont
  {J.}~\bibnamefont {Lesueur}},\ }\bibfield  {title} {\enquote {\bibinfo
  {title} {Two-dimensional superconducting phase in
  $\mathrm{LaAlO}{}_{3}/\mathrm{SrTiO}{}_{3}$ heterostructures induced by
  high-mobility carrier doping},}\ }\href {\doibase
  10.1103/PhysRevLett.108.247004} {\bibfield  {journal} {\bibinfo  {journal}
  {Phys. Rev. Lett.}\ }\textbf {\bibinfo {volume} {108}},\ \bibinfo {pages}
  {247004} (\bibinfo {year} {2012})}\BibitemShut {NoStop}%
\bibitem [{\citenamefont {Diez}\ \emph {et~al.}(2015)\citenamefont {Diez},
  \citenamefont {Monteiro}, \citenamefont {Mattoni}, \citenamefont {Cobanera},
  \citenamefont {Hyart}, \citenamefont {Mulazimoglu}, \citenamefont {Bovenzi},
  \citenamefont {Beenakker},\ and\ \citenamefont {Caviglia}}]{Diez2015}%
  \BibitemOpen
  \bibfield  {author} {\bibinfo {author} {\bibfnamefont {M.}~\bibnamefont
  {Diez}}, \bibinfo {author} {\bibfnamefont {A.~M. R. V.~L.}\ \bibnamefont
  {Monteiro}}, \bibinfo {author} {\bibfnamefont {G.}~\bibnamefont {Mattoni}},
  \bibinfo {author} {\bibfnamefont {E.}~\bibnamefont {Cobanera}}, \bibinfo
  {author} {\bibfnamefont {T.}~\bibnamefont {Hyart}}, \bibinfo {author}
  {\bibfnamefont {E.}~\bibnamefont {Mulazimoglu}}, \bibinfo {author}
  {\bibfnamefont {N.}~\bibnamefont {Bovenzi}}, \bibinfo {author} {\bibfnamefont
  {C.~W.~J.}\ \bibnamefont {Beenakker}}, \ and\ \bibinfo {author}
  {\bibfnamefont {A.~D.}\ \bibnamefont {Caviglia}},\ }\bibfield  {title}
  {\enquote {\bibinfo {title} {Giant negative magnetoresistance driven by
  spin-orbit coupling at the $\mathrm{LaAlO}{}_{3}/\mathrm{SrTiO}{}_{3}$
  interface},}\ }\href {\doibase 10.1103/PhysRevLett.115.016803} {\bibfield
  {journal} {\bibinfo  {journal} {Phys. Rev. Lett.}\ }\textbf {\bibinfo
  {volume} {115}},\ \bibinfo {pages} {016803} (\bibinfo {year}
  {2015})}\BibitemShut {NoStop}%
\bibitem [{\citenamefont {Khalsa}\ \emph {et~al.}(2013)\citenamefont {Khalsa},
  \citenamefont {Lee},\ and\ \citenamefont {MacDonald}}]{Khalsa2013}%
  \BibitemOpen
  \bibfield  {author} {\bibinfo {author} {\bibfnamefont {G.}~\bibnamefont
  {Khalsa}}, \bibinfo {author} {\bibfnamefont {B.}~\bibnamefont {Lee}}, \ and\
  \bibinfo {author} {\bibfnamefont {A.~H.}\ \bibnamefont {MacDonald}},\
  }\bibfield  {title} {\enquote {\bibinfo {title} {Theory of ${t}_{2g}$
  electron-gas rashba interactions},}\ }\href {\doibase
  10.1103/PhysRevB.88.041302} {\bibfield  {journal} {\bibinfo  {journal} {Phys.
  Rev. B}\ }\textbf {\bibinfo {volume} {88}},\ \bibinfo {pages} {041302}
  (\bibinfo {year} {2013})}\BibitemShut {NoStop}%
\bibitem [{\citenamefont {van Heeringen}\ \emph {et~al.}(2013)\citenamefont
  {van Heeringen}, \citenamefont {de~Wijs}, \citenamefont {McCollam},
  \citenamefont {Maan},\ and\ \citenamefont {Fasolino}}]{Heeringen2013}%
  \BibitemOpen
  \bibfield  {author} {\bibinfo {author} {\bibfnamefont {L.~W.}\ \bibnamefont
  {van Heeringen}}, \bibinfo {author} {\bibfnamefont {G.~A.}\ \bibnamefont
  {de~Wijs}}, \bibinfo {author} {\bibfnamefont {A.}~\bibnamefont {McCollam}},
  \bibinfo {author} {\bibfnamefont {J.~C.}\ \bibnamefont {Maan}}, \ and\
  \bibinfo {author} {\bibfnamefont {A.}~\bibnamefont {Fasolino}},\ }\bibfield
  {title} {\enquote {\bibinfo {title}
  {k$\ifmmode\cdot\else\textperiodcentered\fi{}$p subband structure of the
  $\mathrm{LaAlO}{}_{3}/\mathrm{SrTiO}{}_{3}$ interface},}\ }\href {\doibase
  10.1103/PhysRevB.88.205140} {\bibfield  {journal} {\bibinfo  {journal} {Phys.
  Rev. B}\ }\textbf {\bibinfo {volume} {88}},\ \bibinfo {pages} {205140}
  (\bibinfo {year} {2013})}\BibitemShut {NoStop}%
\bibitem [{\citenamefont {W\'ojcik}\ \emph {et~al.}(2021)\citenamefont
  {W\'ojcik}, \citenamefont {Nowak},\ and\ \citenamefont
  {Zegrodnik}}]{Wojcik2021}%
  \BibitemOpen
  \bibfield  {author} {\bibinfo {author} {\bibfnamefont {P.}~\bibnamefont
  {W\'ojcik}}, \bibinfo {author} {\bibfnamefont {M.~P.}\ \bibnamefont {Nowak}},
  \ and\ \bibinfo {author} {\bibfnamefont {M.}~\bibnamefont {Zegrodnik}},\
  }\bibfield  {title} {\enquote {\bibinfo {title} {Impact of spin-orbit
  interaction on the phase diagram and anisotropy of the in-plane critical
  magnetic field at the superconducting
  $\mathrm{LaAlO}{}_{3}/\mathrm{SrTiO}{}_{3}$ interface},}\ }\href {\doibase
  10.1103/PhysRevB.104.174503} {\bibfield  {journal} {\bibinfo  {journal}
  {Phys. Rev. B}\ }\textbf {\bibinfo {volume} {104}},\ \bibinfo {pages}
  {174503} (\bibinfo {year} {2021})}\BibitemShut {NoStop}%
\bibitem [{\citenamefont {Zegrodnik}\ and\ \citenamefont
  {W\'ojcik}(2020)}]{Zegrodnik2020}%
  \BibitemOpen
  \bibfield  {author} {\bibinfo {author} {\bibfnamefont {M.}~\bibnamefont
  {Zegrodnik}}\ and\ \bibinfo {author} {\bibfnamefont {P.}~\bibnamefont
  {W\'ojcik}},\ }\bibfield  {title} {\enquote {\bibinfo {title}
  {Superconducting dome in $\mathrm{LaAlO}{}_{3}/\mathrm{SrTiO}{}_{3}$
  interfaces as a direct consequence of the extended $s$-wave symmetry of the
  gap},}\ }\href {\doibase 10.1103/PhysRevB.102.085420} {\bibfield  {journal}
  {\bibinfo  {journal} {Phys. Rev. B}\ }\textbf {\bibinfo {volume} {102}},\
  \bibinfo {pages} {085420} (\bibinfo {year} {2020})}\BibitemShut {NoStop}%
\bibitem [{\citenamefont {Caviglia}\ \emph {et~al.}(2010)\citenamefont
  {Caviglia}, \citenamefont {Gabay}, \citenamefont {Gariglio}, \citenamefont
  {Reyren}, \citenamefont {Cancellieri},\ and\ \citenamefont
  {Triscone}}]{Caviglia2010}%
  \BibitemOpen
  \bibfield  {author} {\bibinfo {author} {\bibfnamefont {A.~D.}\ \bibnamefont
  {Caviglia}}, \bibinfo {author} {\bibfnamefont {M.}~\bibnamefont {Gabay}},
  \bibinfo {author} {\bibfnamefont {S.}~\bibnamefont {Gariglio}}, \bibinfo
  {author} {\bibfnamefont {N.}~\bibnamefont {Reyren}}, \bibinfo {author}
  {\bibfnamefont {C.}~\bibnamefont {Cancellieri}}, \ and\ \bibinfo {author}
  {\bibfnamefont {J.-M.}\ \bibnamefont {Triscone}},\ }\bibfield  {title}
  {\enquote {\bibinfo {title} {Tunable rashba spin-orbit interaction at oxide
  interfaces},}\ }\href {\doibase 10.1103/PhysRevLett.104.126803} {\bibfield
  {journal} {\bibinfo  {journal} {Phys. Rev. Lett.}\ }\textbf {\bibinfo
  {volume} {104}},\ \bibinfo {pages} {126803} (\bibinfo {year}
  {2010})}\BibitemShut {NoStop}%
\bibitem [{\citenamefont {Yin}\ \emph {et~al.}(2020)\citenamefont {Yin},
  \citenamefont {Seiler}, \citenamefont {Tang}, \citenamefont {Leermakers},
  \citenamefont {Lebedev}, \citenamefont {Zeitler},\ and\ \citenamefont
  {Aarts}}]{Yin2020}%
  \BibitemOpen
  \bibfield  {author} {\bibinfo {author} {\bibfnamefont {Ch.}\ \bibnamefont
  {Yin}}, \bibinfo {author} {\bibfnamefont {P.}~\bibnamefont {Seiler}},
  \bibinfo {author} {\bibfnamefont {L.~M.~K.}\ \bibnamefont {Tang}}, \bibinfo
  {author} {\bibfnamefont {I.}~\bibnamefont {Leermakers}}, \bibinfo {author}
  {\bibfnamefont {N.}~\bibnamefont {Lebedev}}, \bibinfo {author} {\bibfnamefont
  {U.}~\bibnamefont {Zeitler}}, \ and\ \bibinfo {author} {\bibfnamefont
  {J.}~\bibnamefont {Aarts}},\ }\bibfield  {title} {\enquote {\bibinfo {title}
  {Tuning rashba spin-orbit coupling at
  $\mathrm{LaAlO}{}_{3}/\mathrm{SrTiO}{}_{3}$ interfaces by band filling},}\
  }\href {\doibase 10.1103/PhysRevB.101.245114} {\bibfield  {journal} {\bibinfo
   {journal} {Phys. Rev. B}\ }\textbf {\bibinfo {volume} {101}},\ \bibinfo
  {pages} {245114} (\bibinfo {year} {2020})}\BibitemShut {NoStop}%
\bibitem [{\citenamefont {Ohtomo}\ and\ \citenamefont
  {Hwang}(2004)}]{Ohtomo2004}%
  \BibitemOpen
  \bibfield  {author} {\bibinfo {author} {\bibfnamefont {A}~\bibnamefont
  {Ohtomo}}\ and\ \bibinfo {author} {\bibfnamefont {H.~Y}\ \bibnamefont
  {Hwang}},\ }\bibfield  {title} {\enquote {\bibinfo {title} {A high-mobility
  electron gas at the $\mathrm{LaAlO}{}_{3}/\mathrm{SrTiO}{}_{3}$
  heterointerface},}\ }\href@noop {} {\bibfield  {journal} {\bibinfo  {journal}
  {Nature}\ }\textbf {\bibinfo {volume} {427}},\ \bibinfo {pages} {423}
  (\bibinfo {year} {2004})}\BibitemShut {NoStop}%
\bibitem [{\citenamefont {Bj\o{}rlig}\ \emph {et~al.}(2020)\citenamefont
  {Bj\o{}rlig}, \citenamefont {Carrad}, \citenamefont {Prawiroatmodjo},
  \citenamefont {von Soosten}, \citenamefont {Gan}, \citenamefont {Chen},
  \citenamefont {Pryds}, \citenamefont {Paaske},\ and\ \citenamefont
  {Jespersen}}]{Jespersen2020}%
  \BibitemOpen
  \bibfield  {author} {\bibinfo {author} {\bibfnamefont {A.~V.}\ \bibnamefont
  {Bj\o{}rlig}}, \bibinfo {author} {\bibfnamefont {D.~J.}\ \bibnamefont
  {Carrad}}, \bibinfo {author} {\bibfnamefont {G.~E. D.~K.}\ \bibnamefont
  {Prawiroatmodjo}}, \bibinfo {author} {\bibfnamefont {M.}~\bibnamefont {von
  Soosten}}, \bibinfo {author} {\bibfnamefont {Y.}~\bibnamefont {Gan}},
  \bibinfo {author} {\bibfnamefont {Y.}~\bibnamefont {Chen}}, \bibinfo {author}
  {\bibfnamefont {N.}~\bibnamefont {Pryds}}, \bibinfo {author} {\bibfnamefont
  {J.}~\bibnamefont {Paaske}}, \ and\ \bibinfo {author} {\bibfnamefont {T.~S.}\
  \bibnamefont {Jespersen}},\ }\bibfield  {title} {\enquote {\bibinfo {title}
  {$g$-factors in $\mathrm{LaAlO}{}_{3}/\mathrm{SrTiO}{}_{3}$ quantum dots},}\
  }\href {\doibase 10.1103/PhysRevMaterials.4.122001} {\bibfield  {journal}
  {\bibinfo  {journal} {Phys. Rev. Mater.}\ }\textbf {\bibinfo {volume} {4}},\
  \bibinfo {pages} {122001} (\bibinfo {year} {2020})}\BibitemShut {NoStop}%
\bibitem [{\citenamefont {Thierschmann}\ \emph {et~al.}(2018)\citenamefont
  {Thierschmann}, \citenamefont {Mulazimoglu}, \citenamefont {Manca},
  \citenamefont {Goswami}, \citenamefont {Klapwijk},\ and\ \citenamefont
  {Caviglia}}]{Thierschmann2020}%
  \BibitemOpen
  \bibfield  {author} {\bibinfo {author} {\bibfnamefont {H.}~\bibnamefont
  {Thierschmann}}, \bibinfo {author} {\bibfnamefont {E.}~\bibnamefont
  {Mulazimoglu}}, \bibinfo {author} {\bibfnamefont {N.}~\bibnamefont {Manca}},
  \bibinfo {author} {\bibfnamefont {S.}~\bibnamefont {Goswami}}, \bibinfo
  {author} {\bibfnamefont {T.~M.}\ \bibnamefont {Klapwijk}}, \ and\ \bibinfo
  {author} {\bibfnamefont {A.~D.}\ \bibnamefont {Caviglia}},\ }\bibfield
  {title} {\enquote {\bibinfo {title} {Transport regimes of a split gate
  superconducting quantum point contact in the two-dimensional
  $\mathrm{LaAlO}{}_{3}/\mathrm{SrTiO}{}_{3}$ superfluid},}\ }\href@noop {}
  {\bibfield  {journal} {\bibinfo  {journal} {Nat. Commun.}\ }\textbf {\bibinfo
  {volume} {9}},\ \bibinfo {pages} {2276} (\bibinfo {year} {2018})}\BibitemShut
  {NoStop}%
\bibitem [{\citenamefont {Guenevere}\ \emph {et~al.}(2017)\citenamefont
  {Guenevere}, \citenamefont {Leijnse}, \citenamefont {Trier}, \citenamefont
  {Chen}, \citenamefont {Christensen}, \citenamefont {von Soosten},
  \citenamefont {Pryds},\ and\ \citenamefont {Jespersen}}]{Guenevere2017}%
  \BibitemOpen
  \bibfield  {author} {\bibinfo {author} {\bibfnamefont {E.~D. K.~P.}\
  \bibnamefont {Guenevere}}, \bibinfo {author} {\bibfnamefont {M.}~\bibnamefont
  {Leijnse}}, \bibinfo {author} {\bibfnamefont {F.}~\bibnamefont {Trier}},
  \bibinfo {author} {\bibfnamefont {Y.}~\bibnamefont {Chen}}, \bibinfo {author}
  {\bibfnamefont {D.~V.}\ \bibnamefont {Christensen}}, \bibinfo {author}
  {\bibfnamefont {M.}~\bibnamefont {von Soosten}}, \bibinfo {author}
  {\bibfnamefont {N.}~\bibnamefont {Pryds}}, \ and\ \bibinfo {author}
  {\bibfnamefont {T.~S.}\ \bibnamefont {Jespersen}},\ }\bibfield  {title}
  {\enquote {\bibinfo {title} {Transport and excitations in a negative-u
  quantum dot at the $\mathrm{LaAlO}{}_{3}/\mathrm{SrTiO}{}_{3}$ interface},}\
  }\href@noop {} {\bibfield  {journal} {\bibinfo  {journal} {Nat. Commun.}\
  }\textbf {\bibinfo {volume} {8}},\ \bibinfo {pages} {395} (\bibinfo {year}
  {2017})}\BibitemShut {NoStop}%
\bibitem [{\citenamefont {Cen}\ \emph {et~al.}(2009)\citenamefont {Cen},
  \citenamefont {Thiel}, \citenamefont {Mannhart},\ and\ \citenamefont
  {Levy}}]{chj1}%
  \BibitemOpen
  \bibfield  {author} {\bibinfo {author} {\bibfnamefont {C.}~\bibnamefont
  {Cen}}, \bibinfo {author} {\bibfnamefont {S.}~\bibnamefont {Thiel}}, \bibinfo
  {author} {\bibfnamefont {J.}~\bibnamefont {Mannhart}}, \ and\ \bibinfo
  {author} {\bibfnamefont {J.}~\bibnamefont {Levy}},\ }\bibfield  {title}
  {\enquote {\bibinfo {title} {Oxide nanoelectronics on demand},}\ }\href@noop
  {} {\bibfield  {journal} {\bibinfo  {journal} {Science}\ }\textbf {\bibinfo
  {volume} {323}},\ \bibinfo {pages} {1026} (\bibinfo {year}
  {2009})}\BibitemShut {NoStop}%
\bibitem [{\citenamefont {Maniv}\ \emph {et~al.}(2016)\citenamefont {Maniv},
  \citenamefont {Ron}, \citenamefont {Goldstein}, \citenamefont {Palevski},\
  and\ \citenamefont {Dagan}}]{chj2}%
  \BibitemOpen
  \bibfield  {author} {\bibinfo {author} {\bibfnamefont {E.}~\bibnamefont
  {Maniv}}, \bibinfo {author} {\bibfnamefont {A.}~\bibnamefont {Ron}}, \bibinfo
  {author} {\bibfnamefont {M.}~\bibnamefont {Goldstein}}, \bibinfo {author}
  {\bibfnamefont {A.}~\bibnamefont {Palevski}}, \ and\ \bibinfo {author}
  {\bibfnamefont {Y.}~\bibnamefont {Dagan}},\ }\bibfield  {title} {\enquote
  {\bibinfo {title} {Tunneling into a quantum confinement created by a
  single-step nanolithography of conducting oxide interfaces},}\ }\href@noop {}
  {\bibfield  {journal} {\bibinfo  {journal} {Phys. Rev. B}\ }\textbf {\bibinfo
  {volume} {94}},\ \bibinfo {pages} {045120} (\bibinfo {year}
  {2016})}\BibitemShut {NoStop}%
\bibitem [{\citenamefont {Schroer}\ \emph {et~al.}(2011)\citenamefont
  {Schroer}, \citenamefont {Petersson}, \citenamefont {Jung},\ and\
  \citenamefont {Petta}}]{Schroer2011}%
  \BibitemOpen
  \bibfield  {author} {\bibinfo {author} {\bibfnamefont {M.~D.}\ \bibnamefont
  {Schroer}}, \bibinfo {author} {\bibfnamefont {K.~D.}\ \bibnamefont
  {Petersson}}, \bibinfo {author} {\bibfnamefont {M.}~\bibnamefont {Jung}}, \
  and\ \bibinfo {author} {\bibfnamefont {J.~R.}\ \bibnamefont {Petta}},\
  }\bibfield  {title} {\enquote {\bibinfo {title} {Field tuning the $g$ factor
  in inas nanowire double quantum dots},}\ }\href {\doibase
  10.1103/PhysRevLett.107.176811} {\bibfield  {journal} {\bibinfo  {journal}
  {Phys. Rev. Lett.}\ }\textbf {\bibinfo {volume} {107}},\ \bibinfo {pages}
  {176811} (\bibinfo {year} {2011})}\BibitemShut {NoStop}%
\bibitem [{\citenamefont {Koppens}\ \emph {et~al.}(2006)\citenamefont
  {Koppens}, \citenamefont {Buizert}, \citenamefont {Tielrooij}, \citenamefont
  {Vink}, \citenamefont {Nowack}, \citenamefont {Meunier}, \citenamefont
  {Kouwenhoven},\ and\ \citenamefont {Vandersypen}}]{Koppens2006}%
  \BibitemOpen
  \bibfield  {author} {\bibinfo {author} {\bibfnamefont {F.~H.~L.}\
  \bibnamefont {Koppens}}, \bibinfo {author} {\bibfnamefont {C.}~\bibnamefont
  {Buizert}}, \bibinfo {author} {\bibfnamefont {K.~J.}\ \bibnamefont
  {Tielrooij}}, \bibinfo {author} {\bibfnamefont {I.~T.}\ \bibnamefont {Vink}},
  \bibinfo {author} {\bibfnamefont {K.~C.}\ \bibnamefont {Nowack}}, \bibinfo
  {author} {\bibfnamefont {T.}~\bibnamefont {Meunier}}, \bibinfo {author}
  {\bibfnamefont {L.~P.}\ \bibnamefont {Kouwenhoven}}, \ and\ \bibinfo {author}
  {\bibfnamefont {L.~M.~K.}\ \bibnamefont {Vandersypen}},\ }\bibfield  {title}
  {\enquote {\bibinfo {title} {Driven coherent oscillations of a single
  electron spin in a quantum dot},}\ }\href@noop {} {\bibfield  {journal}
  {\bibinfo  {journal} {Nature}\ }\textbf {\bibinfo {volume} {442}},\ \bibinfo
  {pages} {766} (\bibinfo {year} {2006})}\BibitemShut {NoStop}%
\bibitem [{\citenamefont {Koppens}\ \emph {et~al.}(2008)\citenamefont
  {Koppens}, \citenamefont {Nowack},\ and\ \citenamefont
  {Vandersypen}}]{Koppens2008}%
  \BibitemOpen
  \bibfield  {author} {\bibinfo {author} {\bibfnamefont {F.~H.~L.}\
  \bibnamefont {Koppens}}, \bibinfo {author} {\bibfnamefont {K.~C.}\
  \bibnamefont {Nowack}}, \ and\ \bibinfo {author} {\bibfnamefont {L.~M.~K.}\
  \bibnamefont {Vandersypen}},\ }\bibfield  {title} {\enquote {\bibinfo {title}
  {Spin echo of a single electron spin in a quantum dot},}\ }\href {\doibase
  10.1103/PhysRevLett.100.236802} {\bibfield  {journal} {\bibinfo  {journal}
  {Phys. Rev. Lett.}\ }\textbf {\bibinfo {volume} {100}},\ \bibinfo {pages}
  {236802} (\bibinfo {year} {2008})}\BibitemShut {NoStop}%
\bibitem [{\citenamefont {Nadj-Perge}\ \emph {et~al.}(2012)\citenamefont
  {Nadj-Perge}, \citenamefont {Pribiag}, \citenamefont {van~den Berg},
  \citenamefont {Zuo}, \citenamefont {Plissard}, \citenamefont {Bakkers},
  \citenamefont {Frolov},\ and\ \citenamefont
  {Kouwenhoven}}]{Nadj-Perge2012_2}%
  \BibitemOpen
  \bibfield  {author} {\bibinfo {author} {\bibfnamefont {S.}~\bibnamefont
  {Nadj-Perge}}, \bibinfo {author} {\bibfnamefont {V.~S.}\ \bibnamefont
  {Pribiag}}, \bibinfo {author} {\bibfnamefont {J.~W.~G.}\ \bibnamefont
  {van~den Berg}}, \bibinfo {author} {\bibfnamefont {K.}~\bibnamefont {Zuo}},
  \bibinfo {author} {\bibfnamefont {S.~R.}\ \bibnamefont {Plissard}}, \bibinfo
  {author} {\bibfnamefont {E.~P. A.~M.}\ \bibnamefont {Bakkers}}, \bibinfo
  {author} {\bibfnamefont {S.~M.}\ \bibnamefont {Frolov}}, \ and\ \bibinfo
  {author} {\bibfnamefont {L.~P.}\ \bibnamefont {Kouwenhoven}},\ }\bibfield
  {title} {\enquote {\bibinfo {title} {Spectroscopy of spin-orbit quantum bits
  in indium antimonide nanowires},}\ }\href {\doibase
  10.1103/PhysRevLett.108.166801} {\bibfield  {journal} {\bibinfo  {journal}
  {Phys. Rev. Lett.}\ }\textbf {\bibinfo {volume} {108}},\ \bibinfo {pages}
  {166801} (\bibinfo {year} {2012})}\BibitemShut {NoStop}%
\bibitem [{\citenamefont {Golovach}\ \emph {et~al.}(2006)\citenamefont
  {Golovach}, \citenamefont {Borhani},\ and\ \citenamefont
  {Loss}}]{Golovach2006}%
  \BibitemOpen
  \bibfield  {author} {\bibinfo {author} {\bibfnamefont {V.~N.}\ \bibnamefont
  {Golovach}}, \bibinfo {author} {\bibfnamefont {M.}~\bibnamefont {Borhani}}, \
  and\ \bibinfo {author} {\bibfnamefont {D.}~\bibnamefont {Loss}},\ }\bibfield
  {title} {\enquote {\bibinfo {title} {Electric-dipole-induced spin resonance
  in quantum dots},}\ }\href {\doibase 10.1103/PhysRevB.74.165319} {\bibfield
  {journal} {\bibinfo  {journal} {Phys. Rev. B}\ }\textbf {\bibinfo {volume}
  {74}},\ \bibinfo {pages} {165319} (\bibinfo {year} {2006})}\BibitemShut
  {NoStop}%
\bibitem [{\citenamefont {Stepanenko}\ \emph {et~al.}(2012)\citenamefont
  {Stepanenko}, \citenamefont {Rudner}, \citenamefont {Halperin},\ and\
  \citenamefont {Loss}}]{Loss2012}%
  \BibitemOpen
  \bibfield  {author} {\bibinfo {author} {\bibfnamefont {D.}~\bibnamefont
  {Stepanenko}}, \bibinfo {author} {\bibfnamefont {M.}~\bibnamefont {Rudner}},
  \bibinfo {author} {\bibfnamefont {B.~I.}\ \bibnamefont {Halperin}}, \ and\
  \bibinfo {author} {\bibfnamefont {D.}~\bibnamefont {Loss}},\ }\bibfield
  {title} {\enquote {\bibinfo {title} {Singlet-triplet splitting in double
  quantum dots due to spin-orbit and hyperfine interactions},}\ }\href
  {\doibase 10.1103/PhysRevB.85.075416} {\bibfield  {journal} {\bibinfo
  {journal} {Phys. Rev. B}\ }\textbf {\bibinfo {volume} {85}},\ \bibinfo
  {pages} {075416} (\bibinfo {year} {2012})}\BibitemShut {NoStop}%
\bibitem [{\citenamefont {Nowack}\ \emph {et~al.}(2007)\citenamefont {Nowack},
  \citenamefont {Koppens}, \citenamefont {Nazarov},\ and\ \citenamefont
  {Vandersypen}}]{Nowack2007}%
  \BibitemOpen
  \bibfield  {author} {\bibinfo {author} {\bibfnamefont {K.~C.}\ \bibnamefont
  {Nowack}}, \bibinfo {author} {\bibfnamefont {F.~H.~L.}\ \bibnamefont
  {Koppens}}, \bibinfo {author} {\bibfnamefont {Yu.~V.}\ \bibnamefont
  {Nazarov}}, \ and\ \bibinfo {author} {\bibfnamefont {L.~M.~K.}\ \bibnamefont
  {Vandersypen}},\ }\bibfield  {title} {\enquote {\bibinfo {title} {Coherent
  control of a single electron spin with electric fields},}\ }\href@noop {}
  {\bibfield  {journal} {\bibinfo  {journal} {Science}\ }\textbf {\bibinfo
  {volume} {31}},\ \bibinfo {pages} {1430} (\bibinfo {year}
  {2007})}\BibitemShut {NoStop}%
\bibitem [{\citenamefont {Nowak}\ \emph {et~al.}(2012)\citenamefont {Nowak},
  \citenamefont {Szafran},\ and\ \citenamefont {Peeters}}]{Nowak2012}%
  \BibitemOpen
  \bibfield  {author} {\bibinfo {author} {\bibfnamefont {M.~P.}\ \bibnamefont
  {Nowak}}, \bibinfo {author} {\bibfnamefont {B.}~\bibnamefont {Szafran}}, \
  and\ \bibinfo {author} {\bibfnamefont {F.~M.}\ \bibnamefont {Peeters}},\
  }\bibfield  {title} {\enquote {\bibinfo {title} {Resonant harmonic generation
  and collective spin rotations in electrically driven quantum dots},}\ }\href
  {\doibase 10.1103/PhysRevB.86.125428} {\bibfield  {journal} {\bibinfo
  {journal} {Phys. Rev. B}\ }\textbf {\bibinfo {volume} {86}},\ \bibinfo
  {pages} {125428} (\bibinfo {year} {2012})}\BibitemShut {NoStop}%
\bibitem [{\citenamefont {Szafran}\ \emph {et~al.}(2024)\citenamefont
  {Szafran}, \citenamefont {Zegrodnik}, \citenamefont {Nowak}, \citenamefont
  {Citro},\ and\ \citenamefont {W\'ojcik}}]{Szafran2023}%
  \BibitemOpen
  \bibfield  {author} {\bibinfo {author} {\bibfnamefont {B.}~\bibnamefont
  {Szafran}}, \bibinfo {author} {\bibfnamefont {M.}~\bibnamefont {Zegrodnik}},
  \bibinfo {author} {\bibfnamefont {M.~P.}\ \bibnamefont {Nowak}}, \bibinfo
  {author} {\bibfnamefont {R.}~\bibnamefont {Citro}}, \ and\ \bibinfo {author}
  {\bibfnamefont {P.}~\bibnamefont {W\'ojcik}},\ }\bibfield  {title} {\enquote
  {\bibinfo {title} {Electric dipole spin resonance in a single- and
  two-electron quantum dot defined in two-dimensional electron gas at the
  $\mathrm{LaAlO}{}_{3}/\mathrm{SrTiO}{}_{3}$ interface},}\ }\href@noop {}
  {\bibfield  {journal} {\bibinfo  {journal} {Phys. Rev. B}\ }\textbf {\bibinfo
  {volume} {109}},\ \bibinfo {pages} {155306} (\bibinfo {year}
  {2024})}\BibitemShut {NoStop}%
\bibitem [{\citenamefont {Fischer}\ \emph {et~al.}(2009)\citenamefont
  {Fischer}, \citenamefont {Trif}, \citenamefont {Coish},\ and\ \citenamefont
  {Loss}}]{lossreview}%
  \BibitemOpen
  \bibfield  {author} {\bibinfo {author} {\bibfnamefont {J.}~\bibnamefont
  {Fischer}}, \bibinfo {author} {\bibfnamefont {M.}~\bibnamefont {Trif}},
  \bibinfo {author} {\bibfnamefont {W.A.}\ \bibnamefont {Coish}}, \ and\
  \bibinfo {author} {\bibfnamefont {D.}~\bibnamefont {Loss}},\ }\bibfield
  {title} {\enquote {\bibinfo {title} {Spin interactions, relaxation and
  decoherence in quantum dots},}\ }\href@noop {} {\bibfield  {journal}
  {\bibinfo  {journal} {Sol. Stat. Commun}\ }\textbf {\bibinfo {volume}
  {149}},\ \bibinfo {pages} {1443} (\bibinfo {year} {2009})}\BibitemShut
  {NoStop}%
\bibitem [{\citenamefont {Climente}\ \emph {et~al.}(2007)\citenamefont
  {Climente}, \citenamefont {Bertoni}, \citenamefont {Goldoni}, \citenamefont
  {Rontani},\ and\ \citenamefont {Molinari}}]{climente}%
  \BibitemOpen
  \bibfield  {author} {\bibinfo {author} {\bibfnamefont {J.~I.}\ \bibnamefont
  {Climente}}, \bibinfo {author} {\bibfnamefont {A.}~\bibnamefont {Bertoni}},
  \bibinfo {author} {\bibfnamefont {G.}~\bibnamefont {Goldoni}}, \bibinfo
  {author} {\bibfnamefont {M.}~\bibnamefont {Rontani}}, \ and\ \bibinfo
  {author} {\bibfnamefont {E.}~\bibnamefont {Molinari}},\ }\bibfield  {title}
  {\enquote {\bibinfo {title} {Magnetic field dependence of triplet-singlet
  relaxation in quantum dots with spin-orbit coupling},}\ }\href@noop {}
  {\bibfield  {journal} {\bibinfo  {journal} {Phys. Rev. B}\ }\textbf {\bibinfo
  {volume} {75}},\ \bibinfo {pages} {081303(R)} (\bibinfo {year}
  {2007})}\BibitemShut {NoStop}%
\bibitem [{\citenamefont {Liu}\ \emph {et~al.}(2023)\citenamefont {Liu},
  \citenamefont {Zhao}, \citenamefont {Qin}, \citenamefont {Ma}, \citenamefont
  {Lu}, \citenamefont {Liu}, \citenamefont {Li}, \citenamefont {Wei},
  \citenamefont {Cheng},\ and\ \citenamefont {Liu}}]{phononlaosto}%
  \BibitemOpen
  \bibfield  {author} {\bibinfo {author} {\bibfnamefont {X.}~\bibnamefont
  {Liu}}, \bibinfo {author} {\bibfnamefont {T.}~\bibnamefont {Zhao}}, \bibinfo
  {author} {\bibfnamefont {Z.}~\bibnamefont {Qin}}, \bibinfo {author}
  {\bibfnamefont {C}~\bibnamefont {Ma}}, \bibinfo {author} {\bibfnamefont
  {F.}~\bibnamefont {Lu}}, \bibinfo {author} {\bibfnamefont {T.}~\bibnamefont
  {Liu}}, \bibinfo {author} {\bibfnamefont {J.}~\bibnamefont {Li}}, \bibinfo
  {author} {\bibfnamefont {S.-H.}\ \bibnamefont {Wei}}, \bibinfo {author}
  {\bibfnamefont {G.}~\bibnamefont {Cheng}}, \ and\ \bibinfo {author}
  {\bibfnamefont {W.-T.}\ \bibnamefont {Liu}},\ }\bibfield  {title} {\enquote
  {\bibinfo {title} {Nonlinear optical phonon spectroscopy revealing polaronic
  signatures of the $\mathrm{LaAlO}{}_{3}/\mathrm{SrTiO}{}_{3}$ interface},}\
  }\href@noop {} {\bibfield  {journal} {\bibinfo  {journal} {Sci. Adv}\
  }\textbf {\bibinfo {volume} {9}},\ \bibinfo {pages} {eadg7037} (\bibinfo
  {year} {2023})}\BibitemShut {NoStop}%
\bibitem [{\citenamefont {Popovi\ifmmode~\acute{c}\else \'{c}\fi{}}\ \emph
  {et~al.}(2008)\citenamefont {Popovi\ifmmode~\acute{c}\else \'{c}\fi{}},
  \citenamefont {Satpathy},\ and\ \citenamefont {Martin}}]{Popovic2008}%
  \BibitemOpen
  \bibfield  {author} {\bibinfo {author} {\bibfnamefont {Z.~S.}\ \bibnamefont
  {Popovi\ifmmode~\acute{c}\else \'{c}\fi{}}}, \bibinfo {author} {\bibfnamefont
  {S.}~\bibnamefont {Satpathy}}, \ and\ \bibinfo {author} {\bibfnamefont
  {R.~M.}\ \bibnamefont {Martin}},\ }\bibfield  {title} {\enquote {\bibinfo
  {title} {Origin of the two-dimensional electron gas carrier density at the
  $\mathrm{LaAlO}{}_{3}/\mathrm{SrTiO}{}_{3}$ interface},}\ }\href {\doibase
  10.1103/PhysRevLett.101.256801} {\bibfield  {journal} {\bibinfo  {journal}
  {Phys. Rev. Lett.}\ }\textbf {\bibinfo {volume} {101}},\ \bibinfo {pages}
  {256801} (\bibinfo {year} {2008})}\BibitemShut {NoStop}%
\bibitem [{\citenamefont {Ruhman}\ \emph {et~al.}(2014)\citenamefont {Ruhman},
  \citenamefont {Joshua}, \citenamefont {Ilani},\ and\ \citenamefont
  {Altman}}]{Ruhman2014}%
  \BibitemOpen
  \bibfield  {author} {\bibinfo {author} {\bibfnamefont {J.}~\bibnamefont
  {Ruhman}}, \bibinfo {author} {\bibfnamefont {A.}~\bibnamefont {Joshua}},
  \bibinfo {author} {\bibfnamefont {S.}~\bibnamefont {Ilani}}, \ and\ \bibinfo
  {author} {\bibfnamefont {E.}~\bibnamefont {Altman}},\ }\bibfield  {title}
  {\enquote {\bibinfo {title} {Competition between kondo screening and
  magnetism at the $\mathrm{LaAlO}{}_{3}/\mathrm{SrTiO}{}_{3}$ interface},}\
  }\href {\doibase 10.1103/PhysRevB.90.125123} {\bibfield  {journal} {\bibinfo
  {journal} {Phys. Rev. B}\ }\textbf {\bibinfo {volume} {90}},\ \bibinfo
  {pages} {125123} (\bibinfo {year} {2014})}\BibitemShut {NoStop}%
\bibitem [{\citenamefont {Gariglio}\ \emph {et~al.}(2015)\citenamefont
  {Gariglio}, \citenamefont {F\'ete},\ and\ \citenamefont
  {Triscone}}]{Gariglio2015}%
  \BibitemOpen
  \bibfield  {author} {\bibinfo {author} {\bibfnamefont {S.}~\bibnamefont
  {Gariglio}}, \bibinfo {author} {\bibfnamefont {A.}~\bibnamefont {F\'ete}}, \
  and\ \bibinfo {author} {\bibfnamefont {J.-M.}\ \bibnamefont {Triscone}},\
  }\bibfield  {title} {\enquote {\bibinfo {title} {Electron confinement at the
  $\mathrm{LaAlO}{}_{3}/\mathrm{SrTiO}{}_{3}$ interface},}\ }\href@noop {}
  {\bibfield  {journal} {\bibinfo  {journal} {J. Condens. Matter Phys.}\
  }\textbf {\bibinfo {volume} {27}},\ \bibinfo {pages} {283201} (\bibinfo
  {year} {2015})}\BibitemShut {NoStop}%
\bibitem [{\citenamefont {Guerrero-Becerra}\ and\ \citenamefont
  {Rontani}(2014)}]{Guerrero2014}%
  \BibitemOpen
  \bibfield  {author} {\bibinfo {author} {\bibfnamefont {K.~A.}\ \bibnamefont
  {Guerrero-Becerra}}\ and\ \bibinfo {author} {\bibfnamefont {M.}~\bibnamefont
  {Rontani}},\ }\bibfield  {title} {\enquote {\bibinfo {title} {Wigner
  localization in a graphene quantum dot with a mass gap},}\ }\href {\doibase
  10.1103/PhysRevB.90.125446} {\bibfield  {journal} {\bibinfo  {journal} {Phys.
  Rev. B}\ }\textbf {\bibinfo {volume} {90}},\ \bibinfo {pages} {125446}
  (\bibinfo {year} {2014})}\BibitemShut {NoStop}%
\bibitem [{\citenamefont {Nowak}\ and\ \citenamefont
  {Szafran}(2014)}]{NowakSzafran14}%
  \BibitemOpen
  \bibfield  {author} {\bibinfo {author} {\bibfnamefont {M.P.}\ \bibnamefont
  {Nowak}}\ and\ \bibinfo {author} {\bibfnamefont {B.}~\bibnamefont
  {Szafran}},\ }\bibfield  {title} {\enquote {\bibinfo {title} {Spontaneous and
  resonant lifting of the spin blockade in nanowire quantum dots},}\
  }\href@noop {} {\bibfield  {journal} {\bibinfo  {journal} {Phys. Rev. B}\
  }\textbf {\bibinfo {volume} {89}},\ \bibinfo {pages} {205412} (\bibinfo
  {year} {2014})}\BibitemShut {NoStop}%
\bibitem [{\citenamefont {Ivakhnenko}\ \emph {et~al.}(2023)\citenamefont
  {Ivakhnenko}, \citenamefont {Shevchenko},\ and\ \citenamefont {Nori}}]{lzs}%
  \BibitemOpen
  \bibfield  {author} {\bibinfo {author} {\bibfnamefont {O.V.}\ \bibnamefont
  {Ivakhnenko}}, \bibinfo {author} {\bibfnamefont {S.N.}\ \bibnamefont
  {Shevchenko}}, \ and\ \bibinfo {author} {\bibfnamefont {F.}~\bibnamefont
  {Nori}},\ }\bibfield  {title} {\enquote {\bibinfo {title} {Nonadiabatic
  landau-zener-st\"uckelberg-majorana transitions, dynamics, and
  interference},}\ }\href@noop {} {\bibfield  {journal} {\bibinfo  {journal}
  {Phys. Reports}\ }\textbf {\bibinfo {volume} {995}},\ \bibinfo {pages} {1}
  (\bibinfo {year} {2023})}\BibitemShut {NoStop}%
\bibitem [{\citenamefont {Khomitsky}\ \emph {et~al.}(2020)\citenamefont
  {Khomitsky}, \citenamefont {Lavrukhina},\ and\ \citenamefont
  {Sherman}}]{Sherman}%
  \BibitemOpen
  \bibfield  {author} {\bibinfo {author} {\bibfnamefont {D.V.}\ \bibnamefont
  {Khomitsky}}, \bibinfo {author} {\bibfnamefont {E.A.}\ \bibnamefont
  {Lavrukhina}}, \ and\ \bibinfo {author} {\bibfnamefont {E.Ya.}\ \bibnamefont
  {Sherman}},\ }\bibfield  {title} {\enquote {\bibinfo {title} {Spin rotation
  by resonant electric field in few-level quantum dots: Floquet dynamics and
  tunneling},}\ }\href {\doibase 10.1103/PhysRevApplied.14.014090} {\bibfield
  {journal} {\bibinfo  {journal} {Phys. Rev. Appl.}\ }\textbf {\bibinfo
  {volume} {14}},\ \bibinfo {pages} {014090} (\bibinfo {year}
  {2020})}\BibitemShut {NoStop}%
\bibitem [{\citenamefont {Laird}\ \emph {et~al.}(2009)\citenamefont {Laird},
  \citenamefont {Barthel}, \citenamefont {Rashba}, \citenamefont {Marcus},
  \citenamefont {Hanson},\ and\ \citenamefont {Gossard}}]{Laird09}%
  \BibitemOpen
  \bibfield  {author} {\bibinfo {author} {\bibfnamefont {E.A.}\ \bibnamefont
  {Laird}}, \bibinfo {author} {\bibfnamefont {C.}~\bibnamefont {Barthel}},
  \bibinfo {author} {\bibfnamefont {E.I.}\ \bibnamefont {Rashba}}, \bibinfo
  {author} {\bibfnamefont {C.M.}\ \bibnamefont {Marcus}}, \bibinfo {author}
  {\bibfnamefont {M.P.}\ \bibnamefont {Hanson}}, \ and\ \bibinfo {author}
  {\bibfnamefont {A.C.}\ \bibnamefont {Gossard}},\ }\bibfield  {title}
  {\enquote {\bibinfo {title} {A new mechanism of electric dipole spin
  resonance: hyperfine coupling in quantum dots},}\ }\href {\doibase
  10.1088/0268-1242/24/6/064004} {\bibfield  {journal} {\bibinfo  {journal}
  {Semic. Sci. Technol.}\ }\textbf {\bibinfo {volume} {24}},\ \bibinfo {pages}
  {064004} (\bibinfo {year} {2009})}\BibitemShut {NoStop}%
\bibitem [{\citenamefont {Klemt}\ \emph {et~al.}(2023)\citenamefont {Klemt},
  \citenamefont {Elhomsy}, \citenamefont {Nurizzo}, \citenamefont {Hamonic},
  \citenamefont {Martinez}, \citenamefont {Paz}, \citenamefont {Spence},
  \citenamefont {Dartiailh}, \citenamefont {Jadot}, \citenamefont {Chanrion},
  \citenamefont {Thiney}, \citenamefont {Lethiecq}, \citenamefont {Bertrand},
  \citenamefont {Niebojewski}, \citenamefont {Bauerle}, \citenamefont {Vinet},
  \citenamefont {Niquet}, \citenamefont {Meunier},\ and\ \citenamefont
  {Urdampilleta}}]{Klemt}%
  \BibitemOpen
  \bibfield  {author} {\bibinfo {author} {\bibfnamefont {B.}~\bibnamefont
  {Klemt}}, \bibinfo {author} {\bibfnamefont {V.}~\bibnamefont {Elhomsy}},
  \bibinfo {author} {\bibfnamefont {M.}~\bibnamefont {Nurizzo}}, \bibinfo
  {author} {\bibfnamefont {P.}~\bibnamefont {Hamonic}}, \bibinfo {author}
  {\bibfnamefont {B.}~\bibnamefont {Martinez}}, \bibinfo {author}
  {\bibfnamefont {B.~C.}\ \bibnamefont {Paz}}, \bibinfo {author} {\bibfnamefont
  {C.}~\bibnamefont {Spence}}, \bibinfo {author} {\bibfnamefont {M.~C.}\
  \bibnamefont {Dartiailh}}, \bibinfo {author} {\bibfnamefont {B.}~\bibnamefont
  {Jadot}}, \bibinfo {author} {\bibfnamefont {E.}~\bibnamefont {Chanrion}},
  \bibinfo {author} {\bibfnamefont {V.}~\bibnamefont {Thiney}}, \bibinfo
  {author} {\bibfnamefont {R.}~\bibnamefont {Lethiecq}}, \bibinfo {author}
  {\bibfnamefont {B.}~\bibnamefont {Bertrand}}, \bibinfo {author}
  {\bibfnamefont {H.}~\bibnamefont {Niebojewski}}, \bibinfo {author}
  {\bibfnamefont {C.}~\bibnamefont {Bauerle}}, \bibinfo {author} {\bibfnamefont
  {M.}~\bibnamefont {Vinet}}, \bibinfo {author} {\bibfnamefont {Y-M.}\
  \bibnamefont {Niquet}}, \bibinfo {author} {\bibfnamefont {T.}~\bibnamefont
  {Meunier}}, \ and\ \bibinfo {author} {\bibfnamefont {M.}~\bibnamefont
  {Urdampilleta}},\ }\bibfield  {title} {\enquote {\bibinfo {title} {Electrical
  manipulation of a single electron spin in cmos using a micromagnet and
  spin-valley coupling},}\ }\href {\doibase doi.org/10.1038/s41534-023-00776-8}
  {\bibfield  {journal} {\bibinfo  {journal} {npj Quantum Inf.}\ }\textbf
  {\bibinfo {volume} {9}},\ \bibinfo {pages} {107} (\bibinfo {year}
  {2023})}\BibitemShut {NoStop}%
\bibitem [{\citenamefont {Philips}\ \emph {et~al.}(2022)\citenamefont
  {Philips}, \citenamefont {Madzik}, \citenamefont {Amitonov}, \citenamefont
  {de~Snoo}, \citenamefont {Russ}, \citenamefont {Kalhor}, \citenamefont
  {Volk}, \citenamefont {Lawrie}, \citenamefont {Brousse}, \citenamefont
  {Tryputen}, \citenamefont {Wuetz}, \citenamefont {Sammak}, \citenamefont
  {Veldhorst}, \citenamefont {G.},\ and\ \citenamefont
  {Vandersypen}}]{Philips}%
  \BibitemOpen
  \bibfield  {author} {\bibinfo {author} {\bibfnamefont {S.~G.~J.}\
  \bibnamefont {Philips}}, \bibinfo {author} {\bibfnamefont {M.~T.}\
  \bibnamefont {Madzik}}, \bibinfo {author} {\bibfnamefont {S.~V.}\
  \bibnamefont {Amitonov}}, \bibinfo {author} {\bibfnamefont {S.~L.}\
  \bibnamefont {de~Snoo}}, \bibinfo {author} {\bibfnamefont {M.}~\bibnamefont
  {Russ}}, \bibinfo {author} {\bibfnamefont {N.}~\bibnamefont {Kalhor}},
  \bibinfo {author} {\bibfnamefont {C.}~\bibnamefont {Volk}}, \bibinfo {author}
  {\bibfnamefont {W.~I.~L.}\ \bibnamefont {Lawrie}}, \bibinfo {author}
  {\bibfnamefont {D.}~\bibnamefont {Brousse}}, \bibinfo {author} {\bibfnamefont
  {L.}~\bibnamefont {Tryputen}}, \bibinfo {author} {\bibfnamefont {B.~P.}\
  \bibnamefont {Wuetz}}, \bibinfo {author} {\bibfnamefont {A.}~\bibnamefont
  {Sammak}}, \bibinfo {author} {\bibfnamefont {M.}~\bibnamefont {Veldhorst}},
  \bibinfo {author} {\bibfnamefont {Scappucci}\ \bibnamefont {G.}}, \ and\
  \bibinfo {author} {\bibfnamefont {L.~M.~K.}\ \bibnamefont {Vandersypen}},\
  }\bibfield  {title} {\enquote {\bibinfo {title} {Universal control of a
  six-qubit quantum processor in silicon},}\ }\href {\doibase
  doi.org/10.1038/s41586-022-05117-x} {\bibfield  {journal} {\bibinfo
  {journal} {Nature}\ }\textbf {\bibinfo {volume} {609}},\ \bibinfo {pages}
  {919} (\bibinfo {year} {2022})}\BibitemShut {NoStop}%
\end{thebibliography}%


%

\end{document}